\documentclass[12pt,letterpaper]{article}
\usepackage{jheppub}

\usepackage{graphicx}
\usepackage{bbm}
\usepackage{amsmath}
\usepackage{amssymb}
\usepackage{mathtools}
\usepackage{amsthm}

\usepackage{customprelude}

\newenvironment{figurehere}{\begin{trivlist}\item\begin{minipage}{\textwidth}\captionsetup{type=figure}\centering}{\end{minipage}\end{trivlist}}

\newcommand{\M}[1]{\mathcal{M}^{#1}}

\newcommand{\secref}[1]{\S\ref{#1}}

\def\be{\begin{equation}}
\def\ee{\end{equation}}

\title{Branes and symmetries for $\cN=3$ S-folds}

\author{Muldrow Etheredge,$^\diamondsuit$}
\author{Iñaki García Etxebarria,$^\heartsuit$}
\author{Ben Heidenreich,$^\diamondsuit$}
\author{and Sebastian Rauch\,$^\diamondsuit$}

\affiliation{$^\diamondsuit$ Amherst Center for Fundamental Interactions,\\Department of Physics, University of Massachusetts, Amherst, MA 01003 USA}
\affiliation{$^\heartsuit$ Department of Mathematical Sciences, Durham University, Durham, DH1 3LE, United Kingdom}

\abstract{We describe the higher-form and non-invertible symmetries of
  4d $\cN= 3$ S-folds using the brane dynamics of their holographic
  duals. In cases with enhancement to $\cN=4$ supersymmetry, our
  analysis reproduces the known field theory results of Aharony,
  Seiberg and Tachikawa, and is compatible with the effective action
  recently given by Bergman and Hirano. Likewise, for two specific $\cN=3$ theories for which Zafrir has conjectured $\cN=1$
  Lagrangians our results agree with those implied by the Lagrangian description. In all other cases, our results imply novel predictions about the symmetries of the corresponding $\mathcal{N}=3$ field theories.}

\begin{document}

\makeatletter
\let\old@fpheader\@fpheader
\renewcommand{\@fpheader}{\old@fpheader\hfill
\hfill AFCI-T23-01}
\makeatother

\maketitle

\section{Introduction}

In recent years there has been a rapid increase in our understanding
of the structure of symmetries in quantum field theory. One of the
foundational papers of this subject is the analysis by Aharony,
Seiberg and Tachikawa of the possible global forms of $\cN=4$ quantum
field theories \cite{Aharony:2013hda}. Those with
unitary, orthogonal or symplectic gauge groups all admit a
holographically dual description in the large $N$ limit, so it is
natural to look for a holographic description of these global forms as well.

For $\mathcal{N}=4$ theories with Lie algebra $\fsu(N)$,\footnote{Henceforward, we
  refer to the Lie algebra $\fg$ 
 instead of the group $G$ when referencing all possible choices
 of gauge group and discrete $\theta$ angles simultaneously.}
   such a holographic description  was
found by Witten~\cite{Witten:1998wy}. The
essential observation is that there is a topological field theory
(TFT) on $\AdS_5$, and different choices of boundary conditions for
this TFT lead to the different choices of global form in the field
theory. To better
understand this TFT and its implications for the global structure of the boundary theory, we replace $\AdS_5$ with a more general manifold
$\M{5}$ which is asymptotically of the form $\M{4}\times \bR$, with
$\M{4}$ closed, Spin, and, for simplicity, without torsion.

The TFT described in \cite{Witten:1998wy} has two types of extended two-surface
operators, which arise from D1 and F1 branes that wrap surfaces in
$\M{5}$. We denote these operators by $\text{D1}(\Sigma^2)$ and
$\text{F1}(\Xi^2)$, respectively. These two operators do not commute
due to the presence of $N$ units of $F_5$ RR flux on the $S^5$ factor,
and instead satisfy the relation
\begin{align}
\text{D1}(\Sigma^2)\text{F1}(\Xi^2) = e^{2\pi i \Sigma^2\cdot \Xi^2/N} \text{F1}(\Xi^2)\text{D1}(\Sigma^2),\label{eq.introF1D1comms}
\end{align}
see, for instance, \S\ref{sec.Sfolds-N=4-brane-noncommutativity} for a derivation.
Here $\Sigma^2$ and $\Xi^2$ are surfaces on $\M{4}$,
$\Sigma^2\cdot \Xi^2$ is their intersection product in $\M{4}$, and we
are viewing the radial direction, $\bR$, as time so that there is a
Hilbert space associated to $\M{4}$ on which the two operators act.

The commutation relations \eqref{eq.introF1D1comms} are those of a discrete $\bZ_N$ gauge theory. The physical content of such a theory is encoded in the spectrum of operators and their algebra,\footnote{To be more precise, one should use the language of extended TFT, see \cite{Kapustin:2010ta} for a review. However, we will not attempt this degree of precision in this paper.} but it is also possible in this case to give a description in terms of continuous fields \cite{Banks:2010zn,Kapustin:2014gua}. This is achieved by the following action,
\begin{align}
  S_{\bZ_N} = \frac{N}{2\pi i} \int_{\M{5}} \bfB_{\text{F1}}\wedge
  d\bfB_{\text{D1}}\, ,
\end{align}
with $\bfB_{\text{F1}}$ and $\bfB_{\text{D1}}$ $2\pi$ periodic two-form connections on a higher $U(1)$ bundle, or in more correct differential cohomology notation (see \cite{Freed:2000ta,Freed:2006yc,Apruzzi:2021nmk} for reviews),
\begin{align}
  S_{\bZ_N} = 2\pi i N \int_{\M{5}} \diff H_{\text{F1}}\star
  \diff H_{\text{D1}},
\end{align}
with $\diff H_{\text{F1}},\diff H_{\text{D1}}\in\diff H^3(\M{5})$. This
is precisely the coupling one obtains on $\M{5}$ after reducing IIB
supergravity on the $S^5$ factor in the long wavelength limit
\cite{Witten:1998wy}.\footnote{This is true if one ignores the
  possibility of singleton modes. A careful analysis of these can be
  found in \cite{Belov:2004ht} (see also
  \cite{Maldacena:2001ss,Gukov:2004id}).}

The remaining $\cN=4$ theories with a large $N$ limit can be obtained
by taking the near horizon limit of D3 branes atop an orientifold
plane. Depending on the sign of the orientifold, the gauge
algebra on the brane is either $\fso(N)$ or $\fsp(N)$. The structure
of the holographic dual is much more subtle in this case. Many
important aspects of this dual were explained in \cite{Witten:1998xy},
and recently Bergman and Hirano~\cite{Bergman:2022otk} have constructed a topological action
in $\AdS_5$ that reproduces all the $\fso(N)$ and $\fsp(N)$ global structures identified in \cite{Aharony:2013hda} together with their $\SL{2,\mathbb{Z}}$ duality orbits.

 Encouraged by these successes, it is natural to apply the holographic viewpoint to study the global structures of field theories with fewer supersymmetries. In particular, in this paper we will do so for the $\cN=3$ S-folds of
\cite{Garcia-Etxebarria:2015wns,Aharony:2016kai}, which are type IIB backgrounds of the form $\AdS_5 \times S^5 / \mathbb{Z}_k$ where the $\mathbb{Z}_k$ torsion one-cycle ($k=2,3,4$ or $6$) carries a suitable discrete Wilson line for the $\mathbb{Z}_k$ subgroup of $\SL{2,\mathbb{Z}}$. As the dual field theories are all non-Lagrangian for $k>2$, they are much harder to access by purely field theoretic means, so the holographic results we obtain are largely new predictions that remain to be verified on the field theory side of the correspondence. However, in two special cases Zafrir has
proposed $\cN=1$ theories whose $\cN=1$ IR fixed point lives on the
same conformal manifold as the $\cN=3$ theories in question \cite{Zafrir:2020epd},
and in these special cases our results agree with those that follow
from the proposed $\cN=1$ Lagrangians.

Since O3 planes can be viewed as $k=2$ S-folds, the $\cN=4$ theories studied by Bergman and
Hirano~\cite{Bergman:2022otk} also fall within our analysis, and we reproduce their results as an additional consistency check.
We also provide a microscopic derivation of their bulk TFT action,
slightly generalizing it by including the zero-form sector and demonstrating how to derive certain cubic couplings responsible for various mixed 't Hooft anomalies and St\"uckelberg
couplings on the field theory side. We further include a detailed dictionary that maps Wilson and
't Hooft lines to bulk branes.

\bigskip

This paper is organized as follows. In \secref{sec.Sfolds}, we derive
the higher form symmetries of S-folds. In \secref{sec.Sfolds-warmup},
we show which kinds of branes are present in the $k=2$ ($\mathcal{N}=4$) case, and we
generalize to arbitrary $k$ in \secref{sec.Sfolds-kgeneral}. In
\secref{sec.Sfolds-N=3-brane-noncommutativity}, we give a microscopic
derivation of the commutation relations between branes for general
S-folds, and we reproduce the $\cN=4$ results in
\secref{sec.Sfolds-N=4-brane-noncommutativity}. In
\secref{sec.Sfolds-effectiveaction}, we produce a microscopic
derivation of the generalization of the effective bulk TFT action of
\cite{Bergman:2022otk} in the $\cN=4$ case. In \secref{sec:k=2-FW-review} and \secref{sec.Sfolds-FW},
we discuss Freed-Witten anomalies. In
\secref{sec.Sfolds-mixedanomalies} we discuss mixed anomalies and certain
non-invertible symmetries that follow from the existence of the mixed
anomalies. In \secref{sec.k=2}, we explain in detail how to connect
our results for the $k=2$ case to known results in the literature. In
\secref{sec.k=2-Dictionary} we provide a dictionary between bulk
worldsheets and field theory lines. Then, in
\secref{sec.k=2-MutualLocality} we use the commutation relations of
\secref{sec.Sfolds} to derive the known mutual locality relations of
\cite{Aharony:2013hda}. We then show in \secref{sec.k=2-orbits} that
the $\SL{2,\mathbb Z}$ duality webs of \cite{Aharony:2013hda} for
$\cN=4$ theories match the duality webs of the bulk theories. We conclude and discuss future directions in~\S\ref{sec:conclusions}.

\section{Higher form symmetries of S-folds}

\label{sec.Sfolds}

Our target is to understand the symmetries of the $\cN=3$ S-folds
constructed in \cite{Garcia-Etxebarria:2015wns,Aharony:2016kai} (see
also \cite{Ferrara:1998zt} for an earlier construction of the
holographic dual we study below). We will do this by directly
computing commutation relations between branes in the holographic
dual, which is of the form $\M{5}\times (S^5/\bZ_k)$, where we view
$S^5$ as the $\sum_i |z_i|^2=1$ base of $\bC^3$, and the $\bZ_k$
action on $S^5$ is then the one induced from the $\bZ_k$ action on
$\bC^3$:
$(z_1,z_2,z_3)\mapsto (\omega_k z_1, \omega_k z_2, \omega_k z_3)$,
with $\omega_k=\exp(2\pi i / k)$. There is additionally a
$\rho_k\in \SL{2,\bZ}$ duality action, when going around the generator
of $\pi_1(S^5/\bZ_k)=\bZ_k$, given by
\begin{align}
  \rho_2 = \begin{pmatrix}-1 & 0\\0&-1\end{pmatrix}\quad ; \quad
  \rho_3 = \begin{pmatrix}-1 & -1\\1&0\end{pmatrix} \quad ; \quad
  \rho_4 = \begin{pmatrix}0 & -1\\1&0\end{pmatrix}\quad ; \quad
  \rho_6 = \begin{pmatrix}0 & -1\\1&1\end{pmatrix}\, .	
\end{align}
These monodromies can be understood as $\bZ_k$ rotations on the
F-theory torus $T^2$. The values of $k$ are restricted to 2, 3, 4, and
6, so that the rotation is an automorphism of the $T^2$ (for specific
values of $\tau$ whenever $k>2$). The $k=2$ case corresponds to the
$\cN=4$ $\fso$ and $\fsp$ theories, while the cases with $k>2$
preserve only $\cN=3$.

By studying the set of allowed dynamical objects of worldvolume
dimension 0, 1, 2 and 3 on $\M{5}$, and their commutation relations,
we learn about the 3, 2, 1, and 0-form symmetries of the field theory
holographically dual to the S-fold. We first review the sources of
these dynamical objects in the more familiar $k=2$ setting, and then
compute the linking pairing for general $k$. With the linking pairing
we compute the commutation relations of the branes in all of the above
S-folds. We then argue for a bulk effective action for the symmetry
TFT, discuss the anomalies of the theories, and explore the
non-invertibility of the symmetries.

\subsection{\alt{$k=2$}{k=2} allowed brane wrappings, a warm up} \label{sec.Sfolds-warmup}

We now review how to generalise the analysis of brane
non-commutativity leading to~\eqref{eq.introF1D1comms} in the
$\fsu(N)$ case to the $\fso(N)$ and $\fsp(N)$ cases. These cases arise
from placing a stack of D3 branes on top of an O3 plane. The
holographic dual can be obtained by taking the near horizon limit, and
it is given by $\M{5}\times \RP^5$, where in $\RP^5\df S^5/\bZ_2$ the
$\bZ_2$ identifies antipodal points, and additionally acts with
$(-1)^{F_L}\Omega$, due to the orientifold action on the
worldsheet. This can be equivalently described as the $k=2$ S-fold,
since what we call $(-1)^{F_L}\Omega$ in worldsheet language can
alternatively be described as $-1\in SL(2,\bZ)$.

The symmetry operators arise from branes wrapping various cycles in the internal space. Branes which are insensitive to the $-1\in \SL{2,\bZ}$ action wrap cycles classified by homology classes in
\begin{align}
	 H_*(\RP^5;\bZ) = \{\bZ, \bZ_2, 0, \bZ_2, 0, \bZ\}\, .
\end{align}
This classification is in particular relevant for D3 branes, which are
indeed singlets under $SL(2,\bZ)$. By wrapping a brane on
$H_0(\RP^5;\bZ)=\bZ$, we obtain domain walls that change the rank of
the theory by one (for instance interpolating from $\fso(2N)$ to
$\fso(2N+2)$). As this changes the local dynamics of the boundary theory, 
we will not consider these walls any further. More relevant to our analysis are branes wrapped on
$H_1(\RP^5;\bZ)=\bZ_2$ and $H_3(\RP^5;\bZ)=\bZ_2$, which lead to dynamical three-surfaces and dynamical strings on
$\M{5}$, respectively. When these objects are pushed to the boundary,
 depending on the choice of boundary conditions
\cite{Gaiotto:2014kfa,Aharony:2016kai} they either become trivial or give
rise to symmetry operators\footnote{To heuristically explain how the insertion of a brane---with its non-trivial \emph{local} degrees of freedom---can give rise to a \emph{topological} defect in the dual CFT, note that as we push the brane towards the boundary we push the corresponding operator insertion farther into the UV. This suppresses any non-topological (dimension larger than 0) piece of the insertion, yielding a topological defect. By contrast, a brane \emph{ending on} the boundary creates a scale-invariant/-covariant defect that is typically not topological.} for 0-form symmetries or 2-form symmetries,
respectively.\footnote{In the 0-form symmetry sector there are
  additional symmetries of a geometric origin---such as the $R$-symmetry group---which we do not
  consider here.}

We will see below that these two kinds of D3 branes do not commute, so
we cannot choose both symmetries to be realized simultaneously
\cite{GarciaEtxebarria:2019caf}: we either have 0-form symmetries on
the boundary, or we have 2-form symmetries. In the simplest cases the
choice comes from the fact that for a given set of local dynamics, we
have the choice of whether to gauge the 0-form symmetry or not. When
we do gauge it, we obtain a ``magnetic'', or ``quantum'', 2-form
symmetry, which when gauged gives back the original 0-form
symmetry. The general situation is complicated by the presence of
mixed 't Hooft anomalies between the 0-form and the 1-form symmetries
for some choices of global form. Gauging the 1-form symmetries leads
to the 0-form symmetries becoming non-invertible
\cite{Choi:2021kmx,Kaidi:2021xfk}. A detailed field theory analysis of
the non-invertible symmetries for the cases discussed in this paper
has been presented in \cite{Bhardwaj:2022yxj} and has been given a
holographic interpretation in \cite{GarciaEtxebarria:2022vzq} (see
also \cite{Apruzzi:2022rei,Heckman:2022muc} for related work in other
cases).

The other set of branes that play a role in this paper are D1/F1
branes, and D5/NS5 branes. The $-1\in SL(2,\bZ)$ monodromy acts
non-trivially on these branes, so their possible wrappings in the
internal space are classified by homology groups with local
coefficients \cite{Hatcher}, which in this case are
\cite{Witten:1998xy}
\begin{align}
  \label{eq.twisted-homology}
  H_*(\RP^5; \tbZ) = \{\bZ_2, 0, \bZ_2, 0, \bZ_2, 0\}\, .
\end{align}
Below we write $\widetilde{\pt}$ for any point in $\RP^5$ when we wish
to view it as a representative of $H_0(\RP^5;\tbZ)=\bZ_2$, and we
choose any $\RP^2\subset \RP^5$ and $\RP^4\subset \RP^5$ as generators
of $H_2(\RP^5;\tbZ)$ and $H_4(\RP^5;\tbZ)$.

We have various possibilities: the D1s and F1s on $\widetilde{\pt}$ give rise to 2-surface excitations on $\M{5}$, which when pushed to the boundary can (again depending on the choice of boundary conditions) give rise to 1-form symmetry generators in the field theory.

We can also wrap 5-branes on $\RP^2$. These give rise to codimension
one objects in $\M{5}$, which lead to domain walls interpolating
between $\fso(2N)$ and $\fso(2N+1)$ (for the D5) or between $\fso(2N)$
or $\fso(2N+1)$ and $\fsp(2N)$ (for the NS5) \cite{Witten:1998xy}. We
will not consider these cases further. Finally, we can wrap 5-branes
on $\RP^4$. Depending on the choice of boundary conditions, these can
lead to symmetry generators for 1-form symmetries when pushed to the
boundary.

\subsection{Commutation relations from topology}

Our goal is to compute the commutation relations of the various bulk
branes in a way that readily generalizes to $\cN=3$ S-folds. We will
first compute the commutation relations in a simplified model, and
then generalize the result to the settings of interest.

Let us ignore for a moment the Chern-Simons terms in IIB supergravity, and consider the generalized Maxwell theory in $d=10$ for the RR field $C_2$
\begin{align}
	  S = \frac{1}{g^2}\int_{\M{5}\times \RP^5} dC_2\wedge * dC_2\, .
\end{align}
In this expression $dC_2$ represents the curvature of an element
$\diff F_3 \in \diff H^3_\tbZ(\M{5}\times \RP^5)$. The $\tbZ$ subindex
indicates that we are not dealing with ordinary cohomology, but rather
cohomology with local coefficients, or ``twisted''
cohomology.\footnote{See \S3.H in \cite{Hatcher} for the relevant
  mathematical background on cohomology with local coefficients, and
  \cite{Freed:2000ta} for how to construct differential cohomology for
  generalized cohomology theories. The cohomology with local
  coefficients that we are using here is a particularly simple
  generalization, since via the F-theory/M-theory duality map (or
  mathematically, via the Leray-Serre spectral sequence) it can be
  understood as the ordinary differential cohomology of a simple
  elliptic fibration over $\RP^5\times \M{5}$.} This is due to the
intrinsic effect of the orientifold action
$(-1)^{F_L}\Omega=-1\in SL(2,\bZ)$ on $\diff F_3$
\cite{Witten:1998xy}.

\paragraph{D5/D1 commutation relations before orientifolding.}

We proceed as in \cite{Moore:2004jv,Freed:2006ya,Freed:2006yc} (the
following is mostly a review of the results in those papers, we refer
the reader to them for more in-depth explanations). We will consider
the \emph{untwisted} classical theory first. This is because the
classical twisted Maxwell theory in our case is rather vacuous: due to
the orientifold projection the $C_2$ field is projected out to
torsional data, so the classical field theory --- which is formulated
in terms of continuous differential forms --- is trivial. On the other
hand, the classical untwisted theory is interesting, and it helps
understand the expressions that arise in the twisted quantum
theory. In the classical untwisted generalized Maxwell theory we can
measure the electric charge by integrating the electric field strength
$* F_3$ over 7-cycles $\Sigma^7$ on $\M{4}\times \RP^5$:\footnote{We
  note that to compute the commutation relations we are using
  canonical quantization, which relies on foliating spacetime with
  Cauchy surfaces. In our Euclidean setting we may take our foliation
  to be along the radial direction of $\M{5}$ near the asymptotic
  boundary. Labeling the field theory spacetime as $\M{4}$, this means
  we are performing our foliation in the asymptotic neighborhood
  $\M{4}\times[0,1]\times\RP^5$. The operators whose commutations we
  are computing are then taken to live in a single leaf
  $\M{4}\times\RP^5$.}
\begin{align}
  q^{\text{classical}}(\Sigma^7) = \int_{\Sigma^7} *F_3\, .
\end{align}
Since $d*F_3=0$, we can equivalently formulate this as
\begin{align}
  q^{\text{classical}}(\varsigma_2) = \int_{\M{4}\times\RP^5}
  *F_3\wedge \varsigma_2\, ,
\end{align}
where $\varsigma_2$ is a representative of the Poincar\'e dual (on $\M{4}\times \RP^5$) to $\Sigma^7$.

The (still untwisted) quantum theory involves a number of modifications. First, already semi-classically we need to make a precise choice of gauge group, since the basic object is the connection. We will make the standard choice that we are in a $U(1)$ theory, so that $C_2$ is a connection on an integrally quantized $U(1)$ 2-bundle. This quantization is automatically encoded in the formalism if we think in terms of a differential cohomology class $\diff F_3$.

We also wish to promote the classical observables to operators acting on states. We can parametrize our state in terms of $\diff F_3$, and we write $\Psi(\diff F_3)$. Because the charges are integrally quantized in the quantum theory, it is more natural to consider the exponentiated charge operator:
\begin{align}
  U_\alpha(\varsigma_2) = \exp\biggl(2\pi i \int_{\M{4}\times\RP^5}
    *F_3\wedge (\alpha \varsigma_2)\biggr)
\end{align}
where $\alpha\in\bT\df\bR/\bZ$. Another standard modification
introduced by the quantum theory is that, since $*F_3$ is canonically
conjugate to $C_2$, canonical quantization implies that
$U_\alpha(\varsigma_2)$ acts by a shift of the wavefunction by
$\alpha\varsigma_2$:
\begin{align}
  U_\alpha(\varsigma_2)\Psi(\diff F_3) = \Psi(\diff
  F_3+i(\alpha\varsigma_2))\, ,
\end{align}
where we have used the inclusion map
$i\colon H^{d-1}(\M{4}\times\RP^5;\bT)\to \diff
H^{d}(\M{4}\times\RP^5)$ of flat connections into differential cohomology.

In fact, as observed in \cite{Freed:2006ya,Freed:2006yc}, in the
quantum theory one needs to generalize this class of operators
slightly to account for torsional effects. The generalization is very
natural: $\varsigma_2\in H^2(\M{4}\times \RP^5;\bZ)$, and
$\alpha\in \bT$, so $\alpha\varsigma_2$ is an element of
$H^2(\M{4} \times\RP^5;\bT)$. But crucially, not every element of
$H^2(\M{4}\times\RP^5;\bT)$ is of this form. Rather, we have a short
exact sequence\footnote{This follows from the universal coefficient
  theorem for cohomology (theorem 2.33 in \cite{DavisAndKirk})
\begin{equation*}
  0 \to H^{n}(X;\bZ)\otimes R \to H^{n}(X;A)\to \Tor_1^{\bZ}(H^{n+1}(X;\bZ), R) \to 0
\end{equation*}
with $R=\bT$ using that $\Tor_1^\bZ(A, B)=\Tor_1^\bZ(\Tor(A), B)$ \cite{Hatcher} and therefore $\Tor_1^\bZ(\bT, A)=\Tor_1^\bZ(\bQ/\bZ, A)=\Tor(A)$, since $\Tor(\bT)=\bQ/\bZ$ and $\Tor_1^\bZ(\bQ/\bZ, A)=\Tor(A)$ \cite{weibel_1994}.}
\begin{align}
  \label{eq.U(1)-cohomology-SES}
  0 \to H^2(\M{4}\times\RP^5;\bR)\otimes \bT \xrightarrow{\theta}
  H^2(\M{4}\times\RP^5;\bT) \xrightarrow{\beta} \Tor
  H^3(\M{4}\times\RP^5;\bZ)\to 0
\end{align}
where $\beta$ is the Bockstein map. In the presence of torsion we can
therefore extend the set of charge operators to $U(\sigma_2)$, with
$\sigma_2$ any flat connection, or equivalently an arbitrary element
of $H^2(\M{4}\times\RP^5;\bT)$. These operators act, by definition, as
\begin{align}
  U(\sigma_2)\Psi(\diff F_3) = \Psi(\diff
  F_3+i(\sigma_2))\, ,
\end{align}
and are interpreted as the operators measuring both integral and torsional electric charges.

We are now ready to compute the desired commutation relations. Consider an operator
\begin{align}
  V(\Xi^2) = \exp\biggl(2\pi i \int_{\Xi^2\times \pt}\diff F_3
  \biggr)
\end{align}
measuring the holonomy of the RR 2-form $C_2$ on $\Xi^2\times\pt\subset \M{4}\times\RP^5$. We have
\begin{align}
  U(\sigma_2)^{-1}V(\Xi^2) U(\sigma_2) = \exp\biggl(2\pi i
    \int_{\Xi^2\times\pt}i(\sigma_2)\biggr) V(\Xi^2)\, .
\end{align}

The phase $\int_{\Xi^2\times\pt}i(\sigma_2)$ is purely topological information descending from the cycles the branes are wrapped on called the linking pairing. In the next section we explicitly compute the linking pairing for the S-folds, allowing us to immediately determine the commutation relations between branes from generalized Maxwell theory. We will see in \S\ref{sec.Sfolds-N=4-brane-noncommutativity} how to treat the commutation relations of branes linked via Chern-Simons terms (such as D1 and F1).

Before we do that, let us show how to compute string/5-brane
commutation relations in the $k=2$ S-fold.

\paragraph{D5/D1 commutation relations after orientifolding.}

We want to detect any non-commutativity of the D1 wrapping a
two-surface $\Sigma^2\times\widetilde{\pt}\subset\M{4}\times \RP^5$,
and a D5 wrapping $\Xi^2\times \RP^4\subset \M{4}\times\RP^5$. More
accurately, what we are computing are the commutation relations of
operators acting on the boundary conditions in the holographic
setup. The non-commutativity comes from couplings of the branes to the
RR fields in the Wess-Zumino terms in the action. We note that we are interpreting the asymptotic D1 and D5 as the symmetry operators themselves (associated to the integrals of $\diff F_3\sim dC_2$ and $*\diff F_3\sim*dC_2$ respectively) \cite{Apruzzi:2022rei,GarciaEtxebarria:2022vzq,Heckman:2022muc}. Choosing $C_2$ as
our basic variable, as above, and still modeling IIB by generalized
Maxwell (we will refine this momentarily), our previous discussion
implies that
\begin{align}
  \text{D5}(\Sigma^2)^{-1}\text{D1}(\Xi^2)\text{D5}(\Sigma^2) =
  \exp\biggl(2\pi i
    \int_{\Xi^2\times\widetilde{\pt}}i(\sigma_2)\biggr)\text{D1}(\Xi^2)
\end{align}
with
$\sigma_2=\beta^{-1}(\PD[\Sigma^2\times\RP^4])=\PD_{\M{4}}[\Sigma^2]\smile
\beta^{-1}(t_1)$, using that this element is torsional, and the
surjective Bockstein $\beta$ in \eqref{eq.U(1)-cohomology-SES}. (We
abuse notation slightly and also denote by $\beta$ the Bockstein
$\beta\colon H^0(\RP^5;\tilde\bT)\to \Tor H^1(\RP^5;\tbZ)$.) Using
$\PD_{\RP^5}[\widetilde{\pt}]=t_5$ (the generator of
$H^5(\RP^5; \tbZ)=\bZ_2$), we have
\begin{align}
  \int_{\widetilde{\pt}} i(\beta^{-1}(t_1)) = i\int_{\RP^5}
  \beta^{-1}(t_1)\smile t_5 = \frac{1}{2}\in \bT\, ,
\end{align}
so
\begin{align}
  \text{D5}(\Sigma^2)^{-1}\text{D1}(\Xi^2)\text{D5}(\Sigma^2) =
  (-1)^{\Xi^2\cdot \Sigma^2}\text{D1}(\Xi^2)\, .
\end{align}
The same arguments apply to the F1/NS5 commutations relations: a F1 wrapping $\Xi^2\times\widetilde{\pt}$ and a NS5 wrapping $\Sigma^2\times \RP^4$ do not commute:
\begin{align}
  \text{NS5}(\Sigma^2)^{-1}\text{F1}(\Xi^2)\text{NS5}(\Sigma^2) =
  (-1)^{\Xi^2\cdot \Sigma^2}\text{F1}(\Xi^2)\, .
\end{align}

\subsection{The branes of general \alt{$k$}{k} and the linking pairing}\label{sec.Sfolds-kgeneral}

The classification of the symmetry operators for general $k$ goes
along the same lines as above, but now the relevant twisted
(co)homology groups are slightly more involved. For a given $k$, the
cohomology groups classifying fields which transform as a
doublet\footnote{We are interested in doublets of $SL(2,\bZ)$ because,
  for $k>2$, the action of $\rho_k$ acts nontrivially within a given
  doublet. Note that $\rho_{k>3}\supset S$ and so mixes e.g. the NS5
  and D5 branes.}  of $SL(2,\bZ)$ are
$H^*(S^5/\bZ_k; (\bZ\oplus\bZ)_{\rho_k})$. For instance, the different
possibilities for introducing 3-form flux are classified by
$H^3(S^5/\bZ_k; (\bZ\oplus\bZ)_{\rho_k})$. This group was computed
\cite{Aharony:2016kai}, using methods that we now review, and which
also allow us to compute all the other (co)homology groups we need for
our analysis (as also done recently in \cite{Heckman:2022muc}). In
general, $H^*(S^5/\bZ_k;A)$ with $A$ a $\bZ_k$-module can be computed
as the homology of the chain complex \cite{Aharony:2016kai}
\begin{align}
  \label{eq.AT-chain-complex}
  C^0 \xrightarrow{1-t} C^1
  \xrightarrow{1+t+t^2+\ldots+t^{k-1}} C^2
  \xrightarrow{1-t} C^3
  \xrightarrow{1+t+t^2+\ldots+t^{k-1}} C^4
  \xrightarrow{1-t} C^5
\end{align}
where $C^i=A$ for all $i\in \{0,\ldots,5\}$, and $t$ the action of $\bZ_k$ on $A$. In the twisted case we have $A=(\bZ\oplus\bZ)_{\rho_k}$ (this is simply $\bZ\oplus\bZ$ seen as a $\bZ_k$ module, with $\rho_k$ the $\bZ_k$ action) and $t=\rho_k$. In the untwisted case $A=\bZ$ and $t=\id$. The differentials alternate between $1+t+t^2+\ldots+t^{k-1}$ and $1-t$, so the composition of two consecutive differentials is $1-t^k=0$. Using that $1+\rho_k+\ldots+\rho_k^{k-1}=0$, and that $\ker(1-\rho_k)=0$, it is immediate to compute
\begin{align}
  H^*(S^5/\bZ_k;(\bZ\oplus\bZ)_{\rho_k}) = \{0, \sC_k, 0, \sC_k, 0,
  \sC_k\}\, ,
\end{align}
where
\begin{align}
  \sC_k\df \coker(1-\rho_k) = \begin{cases}
    \bZ_2\oplus\bZ_2 & \text{for } k=2\, ,\\
    \bZ_3 & \text{for } k=3\, ,\\
    \bZ_2 & \text{for } k=4\, ,\\
    \bZ_1 & \text{for } k=6\, .
  \end{cases}
\end{align}
The twisted homology groups follow from here by Poincar\'e duality:
\begin{align}
  \label{eq.general-twisted-homology}
  H_*(S^5/\bZ_k;(\bZ\oplus\bZ)_{\rho_k}) = \{\sC_k, 0, \sC_k, 0,
  \sC_k, 0\}\, .
\end{align}
These groups determine where we can wrap 1-branes and 5-branes. (Due to the non-trivial monodromy it is not well defined to talk about specific $(p,q)$ charges any more, such as F1s and D1s separately.)

For completeness, we also list here the cohomology groups in the
untwisted case. They can be computed from the formula above taking
$A=\bZ$ and $t=1$, so that $1+t+\ldots+t^{k-1}=k$ and $1-t=0$. We find
\begin{align}
  H^*(S^5/\bZ_k;\bZ) = \{\bZ, 0, \bZ_k, 0, \bZ_k, \bZ\}
\end{align}
and
\begin{align}
  H_*(S^5/\bZ_k;\bZ) = \{\bZ, \bZ_k, 0, \bZ_k, 0, \bZ\}\, .
\end{align}
This case is relevant for the classification of wrapped D3 branes.

The case with no internal fluxes extends the case of $\fso(2n)$
$\cN=4$ theories, where the mixed anomaly (see \S\ref{sec.Sfolds-FW}) plays an
important role. A similar type of anomaly exists in the $k>2$ ($\cN=3$)
cases. To see this, we need to discuss the pairing between elements in
$H^*(S^5/\bZ_k;(\bZ\oplus\bZ)_{\rho_k})$. One convenient way of
computing these is by going to the M-theory dual, as in
\cite{Cvetic:2021sxm,Heckman:2022muc}, where we have ordinary homology
groups with global coefficients, at the cost of introducing an
additional torus fiber.

Consider first the simpler case of the linking pairing on $H_0(S^1/\bZ_k; (\bZ\oplus\bZ)_{\rho_k})$, where the $\bZ_k$ action on $S^1=\bR/\bZ$ is by shifts by $1/k$. Topologically $S^1/\bZ_k=S^1$, but we use this notation to remind ourselves of the non-trivial $SL(2,\bZ)$ holonomy acting on the $\bZ\oplus\bZ$ coefficients. Physically, due to F/M-theory duality, we expect this group to be related to the homology groups of the mapping torus $\sM_k\df ([0,1]\times T^2)/\sim$ where the identification is $(0, z)\sim (1, \omega_k z)$. Here $z$ is a complex coordinate for the $T^2$, and $\omega_k = \exp(2\pi i/k)$. This gluing is only possible for $k=\{1,2,3,4,6\}$. For $k>2$ the complex structure is restricted: the gluing can only be done consistently for $\tau=\exp(2\pi i/k)$.

Mathematically, the connection goes via the Leray-Serre spectral sequence:\footnote{See~\cite{McCleary} for a nice introduction to   spectral sequences.} for any fibration $F\to X\to B$ this is a spectral sequence with second page
\begin{align}
  \label{eq.Leray-Serre}
  E^2_{p,q} = H_p(B; H_q(F;\bZ))
\end{align}
abutting to $H_{p+q}(X)$. Note that the coefficient system in~\eqref{eq.Leray-Serre} is a local one. In our case $B=S^1/\bZ_k$ and $F=T^2$, and the only non-vanishing entries in the second page are
\begin{align}
  E^2_{0,0} =   E^2_{0,2} = E^2_{1,0} = E^2_{1,2} = \bZ \qquad ; \qquad E^2_{0,1} = \sC_k\, .
\end{align}
Here we use that the holonomy acts trivially on $H_0(T^2;\bZ)=\bZ$ and $H_2(T^2;\bZ)=\bZ$, and acts via $\rho_k$ on $H_1(T^2;\bZ)=\bZ\oplus\bZ$, so that a discussion similar to the one around~\eqref{eq.AT-chain-complex} applies. All differentials vanish for degree reasons, so the spectral sequence converges to
\begin{align}
  E^\infty_{0,0} = E^\infty_{0,2} = E^\infty_{1,0} = E^\infty_{1,2} =
  \bZ \qquad ; \qquad E^\infty_{0,1} = \sC_k\, .
\end{align}
The only possibly non-trivial extension comes from the filtration of $H_1(\sM_k;\bZ)$, where we have
\begin{align}
  0 \subset F_0H_1(\sM_k;\bZ) \subset H_1(\sM_k;\bZ)
\end{align}
with
\begin{align}
  \sC_k = E^\infty_{0,1} = \frac{F_0H_1(\sM_k;\bZ)}{F_{-1}H_1(\sM_k;\bZ)=0}\qquad ; \qquad \bZ = E^\infty_{1,0} = \frac{H_1(\sM_k;\bZ)}{F_0H_1(\sM_k;\bZ)}\, .
\end{align}
From here we conclude that $H_1(\sM_k;\bZ)=\bZ\oplus \sC_k$, since $\Ext(\bZ, -)=0$, so gathering results we have (as in \cite{Cvetic:2021sxm,Heckman:2022muc})
\begin{align}
  H_*(\sM_k;\bZ) = \{\bZ, \bZ\oplus\sC_k, \bZ, \bZ\}\, .
\end{align}
One can interpret this result directly from the geometry. Recall that $\sM_k$ is a $T^2$ fibration over a circle, so $H_0(\sM_k;\bZ)=H_3(\sM_k;\bZ)=\bZ$ follows from connectedness and Poincar\'e duality. $H_2(\sM_k;\bZ)=\bZ$ is generated by the class of the $T^2$ fiber, and the $\bZ$ factor in $H_1(\sM_k;\bZ)$ is its Poincar\'e dual, given by the $z=0$ section of the fibration. Our interest is in the remaining $\sC_k$ factor, which comes from fibering a one-cycle in $T^2$ over the base. Due to the non-trivial monodromy for $k>1$ this cycle becomes torsional. Consider for instance the $k=4$ case (which requires $\tau=i$). As our generator, we choose a point $*$ on the base and the $A$ cycle on the $T^2$ over it (we take the standard choice of $A$ and $B$ generators on the $T^2$, given by the horizontal and vertical cycles). Consider the chain $C_1$ obtained by dragging this 1-cycle on the fiber once around the base. Due to the monodromy $\rho_4$, after going around the base $A$ transforms to $B$, so $\partial C_1 = *\times A - (*\times B)$. So in homology $[*\times A]=[*\times B]$. Dragging the cycle twice around the base, the action of $\rho_4^2$ sends $A$ to $-A$, so by the same reasoning $2[*\times A] = 0$. This shows that the homology generated by the $A$ and $B$ cycles on the fiber times a point in the base projects down to $\bZ_2$. More generally the boundary of a chain that goes once around the base identifies any cycle $\gamma$ on $T^2$ with $\rho_k\gamma$, so the part of $H_1(T^2;\bZ)$ that gives non-trivial contributions to $H_1(\sM_k;\bZ)$ is indeed $\sC_k\df\coker(1-\rho_k)$.

\medskip

We are interested in the linking pairing
\begin{align}
  \sL \colon \Tor H_1(\sM_k;\bZ)\times \Tor H_1(\sM_k;\bZ)\to \bT\, ,
\end{align}
where again $\bT\df \bR/\bZ$ (thought of as an abelian group with
addition). Recall that this torsional pairing is defined as
follows. Given two torsional cycles $a,b\in \Tor H_1(\sM_k;\bZ)$ there
is some $n\in\bZ$ such that $na=0$ in homology. This implies that
there is a 2-chain $C$ such that $\partial C = na$. We define
$\sL(a,b) = C\cdot b/n$ mod 1. (The result of this computation does
not depend on the possible choices one can make.) In the case at hand,
choose $a$ to be a generator of $\Tor H_1(\sM_k;\bZ)=\sC_k$ given a
1-cycle on $T^2$ times a point $*$ in the base $S^1$. Consider for
instance the case $k=4$. Then we know that we can choose
$a=*\times A$, $n=2$, and $C$ is the total space that arises by
dragging the $A$ cycle over the base twice. The relevant intersections
of $a$ and $C$ are at the point $*$ on the base, once with the cycle
$A$ itself and once (after going around the base $S^1$) with
$B$. Using $A\cdot A = 0$, $\rho_4A=B$ and $A\cdot B=1$, we find
\begin{align}
  \sL(a,a) = \frac{1}{2} (A + \rho_4 A)\cdot A = \frac{1}{2} \mod 1\, .
\end{align}
The $k=2$ case works similarly, to give
\begin{align}
  \sL = \begin{pmatrix}
          0 & \frac{1}{2}\\
          \frac{1}{2} & 0
        \end{pmatrix}
        \mod 1
   \label{eq.L2}
\end{align}
written on the natural $[*\times A]$ and $[*\times B]$ basis. The
$k=3$ case is slightly more subtle. We choose $a=[*\times A]$. Note
that $\rho_3A = -A+B$ and $\rho_3^2A=-B$. So
$3 (*\times A) = \partial (C_1 + C_2)$, where $C_1$ is the total space
of taking the $A$ cycle once around the base, with boundary
$A-\rho_3A=2A-B$ and $C_2$ the total space of taking $A$ twice around
the base, with boundary $A-\rho_3^2A = A+B$. From here
\begin{align}
  \sL(a,a) = \frac{1}{3}(-A+B)\cdot A = -\frac{1}{3}\mod 1\, ,
\end{align}
in agreement with the result in \cite{Cvetic:2021sxm}. Finally, when $k=6$ the linking pairing is trivial, since $\sC_6=\bZ_1$.

\medskip

We now extend this discussion to $S^5/\bZ_k$. This is in fact fairly straightforward: if we represent $S^5$ as a circle fibration over $\bC\bP^2$, the $\bZ_k$ action acts purely on the fiber as a $1/k$ shift, precisely as above.\footnote{To see this, recall that we can construct $\bC\bP^2$ starting from $\bC^3$ in two steps: first we take the unit radius sphere $\sum_i |z_i|^2=1$ over the origin, and then quotient by the $U(1)$ action $(z_1,z_2,z_3)\to (e^{i\alpha}z_1, e^{i\alpha}z_2, e^{i\alpha}z_3)$. The $U(1)$ quotient gives as a projection map $p\colon S^5\to \bC\bP^2$, with fiber $S^1$. Clearly the $\bZ_k$ action $(z_1, z_2, z_3)\to (\omega_k z_1, \omega_k z_2, \omega_k z_2)$ then acts on the $S^1$ fiber as a shift, leaving the $\bC\bP^2$ base invariant.} The uplift of the non-trivial $SL(2,\bZ)$ bundle to M-theory is therefore a fibration of $\sM_k$ over $\bC\bP^2$.

The origin of the twisted homology
groups~\eqref{eq.general-twisted-homology} in this picture is then
clear: the generators of
$H_{\text{even}}(S^5/\bZ_k;(\bZ\oplus\bZ)_{\rho_k})=\sC_k$ arise from
the generators of $H_1(\sM_k;\bZ)=\sC_k$ fibered over the generators
of $H_*(\bC\bP^2;\bZ)=\{\bZ, 0, \bZ, 0, \bZ\}$, and the linking
pairing is the linking pairing on $\sM_k$ times the intersection
product on the $\bC\bP^2$ base.

\subsection{Commutation relations for \alt{$\cN=3$}{N=3} S-folds}\label{sec.Sfolds-N=3-brane-noncommutativity}

With this information at hand we can work out the commutation
relations between the generators of the higher form symmetries for
S-folds, including the effective theory on $\M{5}$. We initially
assume that all internal fluxes vanish, so we do not need to concern
ourselves with some of the symmetry generators being projected out due
to the Freed-Witten anomaly \cite{Freed:1999vc}. (We will address this
point in \S\ref{sec.Sfolds-FW} below.) The discussion of the $\cN=4$
cases (that is, $k=2$) is more involved than the $\cN=3$ cases
($k>2$), so we will postpone the analysis of the former to
\S\ref{sec.Sfolds-N=4-brane-noncommutativity}.

In this case we have symmetry generators arising from D3 branes wrapping generators $\Gamma^1$ of $H_1(S^5/\bZ_k;\bZ)$ and $\Gamma^3$ of $H_3(S^5/\bZ_k;\bZ)$, associated (depending on the choice of boundary conditions) with $\bZ_k$ 0-form and 2-form symmetries on the boundary theory. These symmetries were the ones described in \cite{Aharony:2016kai}. The commutation relations of these operators on $\M{4}$ are straightforward to compute given the linking pairing for untwisted cohomology (the geometric linking pairing). We have, for $k>2$
\begin{align}
  \text{D3}(\Sigma^3\times \Gamma^1)^{-1} \text{D3}(\Xi^1\times\Gamma^3)\text{D3}(\Sigma^3\times \Gamma^1) = e^{2\pi i\, \Xi^1\cdot\Sigma^3/k} \text{D3}(\Xi^1\times\Gamma^3)
\end{align}

As in $\fsu$, we can alternatively describe the $\bZ_k$ effective theory on $\M{5}$ describing these branes in terms of dynamical $\bfA_1$ and $\bfA_3$ fields with an effective action
\begin{align}
  S_A^{(k)} = 2\pi i k\int_{\M{5}} \bfA_1\wedge d\bfA_3\, .
\end{align}

We also have 1-form symmetries arising from wrapping 5-branes and
1-branes on cycles in the internal space. These branes transform
non-trivially under $SL(2,\bZ)$, so the different wrapping
possibilities are those classified
by~\eqref{eq.general-twisted-homology}. We can wrap 5-branes on
$H_2(S^5/\bZ_k;(\bZ\oplus\bZ)_{\rho_k})=\sC_k$ and
$H_4(S^5/\bZ_k;(\bZ\oplus\bZ)_{\rho_k})=\sC_k$. The first possibility
leads to generators of $(-1)$-form symmetries, which we will not
discuss further in this paper. The second possibility leads to a
generator of a 1-form $\sC_k$ symmetry. Similarly we can wrap 1-branes
on $H_0(S^5/\bZ_k;(\bZ\oplus\bZ)_{\rho_k})=\sC_k$ and
$H_2(S^5/\bZ_k;(\bZ\oplus\bZ)_{\rho_k})=\sC_k$, leading to generators
of 1-form and 3-form symmetries. We again focus on the 1-form symmetry
part. The commutator between the 1-brane and the 5-brane symmetry
generators, which we will denote as $\bI$ and $\bV$ respectively, is,
for $k>2$
\begin{align}
  \bI(\Sigma^2)^{-1} \bV(\Xi^2) \bI(\Sigma^2) = \exp\biggl(\frac{2\pi
      i}{|\sC_k|} \Sigma^2\cdot\Xi^2\biggr) \bV(\Xi^2)\, .
\end{align}
We have, in other words, a $\sC_k$ gauge theory, which we can also represent as
\begin{align}
  S_{BF}^{(k)} = 2\pi i |\sC_k| \int_{\M{5}} \bfB_{\bI} \wedge d\bfB_{\bV}
\end{align}
where $\bfB_\bI$ and $\bfB_\bV$ are $U(1)$-valued 2-form connections.

We note explicitly that there are no string/string or 5-brane/5-brane commutation relations because there are not distinct $(p,q)$ configurations, due to the nontrivial monodromy. Since there is only one string state and one 5-brane state they trivially must commute with themselves.

\subsection{Commutation relations for \alt{$\cN=4$}{N=4} S-folds}
\label{sec.Sfolds-N=4-brane-noncommutativity}

This case is somewhat more complicated than the previous one, mainly
because the components of the $\SL{2,\bZ}$ doublet stay distinct. In
particular we will see that the commutation relations between F1/D1
and D5/NS5 are nontrivial to compute and essential to the analysis.

\paragraph{String/5-brane commutation relations}

Given the linking pairing \ref{eq.L2} it is immediate to compute the D1/D5 and F1/NS5 commutation relations as
\begin{equation}
\begin{aligned}
  \text{D5}(\Sigma^2)^{-1}\text{D1}(\Xi^2)\text{D5}(\Sigma^2) =
  (-1)^{\Xi^2\cdot \Sigma^2}\text{D1}(\Xi^2)\, , \\
  \text{NS5}(\Sigma^2)^{-1}\text{F1}(\Xi^2)\text{NS5}(\Sigma^2) =
  (-1)^{\Xi^2\cdot \Sigma^2}\text{F1}(\Xi^2)\, .
\end{aligned}
\end{equation}

So far we have ignored the Chern-Simons terms in the IIB supergravity
action. Given that one of the original examples of brane
non-commutativity \cite{Witten:1998wy} relies on the existence of the
Chern-Simons terms, this is a fairly large omission, which we now
remedy. We will focus on the $C_4\wedge H_3\wedge F_3$ term in the IIB
(pseudo) action, which is the one that plays a fundamental role both
in \cite{Witten:1998wy} and in our analysis. This term affects our
discussion above in that it add a term proportional to $C_4\wedge H_3$
to the canonical momentum conjugate to $C_2$.\footnote{The situation
  is complicated slightly due to the fact that the gauge invariant
  fluxes appearing in the string theory action are of the form
  $\tilde F_3=dC_2-H_3\wedge C_0$. An analysis taking these
  complications into account (most efficiently done using the M-theory
  dual) leads to the same conclusion as in the text.} This does not
modify our conclusions above about the commutator of the D1 and the
D5: if we choose
$\sigma_2=\beta^{-1}(\PD_{\M{4}}[\Sigma^2\times\RP^4])$ as above the
contribution of the new term to $\text{D5}(\Sigma^2)$ is
\begin{align}
  \exp\biggl(2\pi i \int_{\Sigma^2\times \RP^4} \diff F_5\star \diff
    H_3\biggr)
\end{align}
and the integral in the exponential vanishes.

\medskip

We next consider the F1/D1 commutator. This commutator was the crucial one in the $\fsu(N)$ case analysed in \cite{Witten:1998wy}, where it was found that the F1 and D1 branes generically did not commute.  In our case, due to the fact that the $\diff H_3$ and $\diff F_3$ fluxes live in twisted cohomology, the analogous commutator will turn out to vanish. To see why this is the case, we will first reformulate the analysis in \cite{Witten:1998wy} in terms of differential cohomology, and then discuss why the orientifold action forces the commutator to vanish.

\paragraph{F1/D1 commutator in the $\fsu(N)$ theory.} In the untwisted case studied in \cite{Witten:1998wy}, due to the $N$ units of $F_5$ flux in the internal space we have
\begin{align}
  \label{eq.su(N)-F1-WZ}
  \text{F1}(\Sigma_2) \df \exp\biggl(2\pi i \int_{\Sigma_2\times \pt}
    \diff H_3\biggr) = \exp\biggl(\frac{2\pi i}{N}\int_{\Sigma_2\times
      S^5}\diff H_3 \star \diff F_5\biggr)
\end{align}
where $\diff F_5$ is the background RR 5-form field. This suggests looking to the generator $\Pi_{\diff F_3}(\diff \eta)$ of $\diff F_3$ displacements by $\diff\eta\df i(\theta(\PD[\Sigma_2\times S^5]\otimes 1/N))$ (the map $\theta$ was defined in~\eqref{eq.U(1)-cohomology-SES}), which does indeed induce the commutation relation~\eqref{eq.F1D1-su(N)}. Due to the presence of the $\diff H_3\star \diff F_3\star \diff F_5$ Chern-Simons term we have
\begin{align}
  \Pi_{\diff F_3}(\diff \eta) = \exp\biggl(\frac{2\pi
      i}{N}\int_{\Sigma_2\times S^5}\diff H_3 \star \diff F_5\biggr)
  \exp\biggl(\frac{2\pi i}{N}\int_{\Sigma_2\times S^5} \diff
    F_7\biggr)\, .
\end{align}
The second term vanishes in the absence of background $F_7$ flux,\footnote{Note that the flux being measured here is the non-torsional part of $\diff F_7$, so it can consistently be set to 0 throughout.} and the first term is the WZ term in the F1 action, as in~\eqref{eq.su(N)-F1-WZ}. The commutation relation found in \cite{Witten:1998wy} now follows immediately
\begin{align}
  \label{eq.F1D1-su(N)}
  \text{(for $\fsu(N))$}\qquad 
  \text{F1}(\Sigma^2)^{-1}\text{D1}(\Xi^2)\text{F1}(\Sigma^2) = \exp\biggl(\frac{2\pi i}{N} \Sigma^2\cdot \Xi^2\biggr)\text{D1}(\Xi^2)\, .
\end{align}

\paragraph{F1/D1 commutator in the orientifolded theory.}

The orientifold changes this analysis significantly.  The main
difference is that $H^2(\M{4}\times \RP^5;\tbZ)$ is purely torsional,
so $H^2(\M{4}\times\RP^5;\tbZ)\otimes \bT=0$, and therefore
$\diff\eta=0$. So inserting an F1 string on
$\Sigma^2\times\widetilde{\pt}$ does not lead to a shift of
$\diff F_3$ for any value of $N$. Accordingly:
\begin{align}
  \text{F1}(\Sigma^2)^{-1}\text{D1}(\Xi^2)\text{F1}(\Sigma^2) =
  \text{D1}(\Xi^2)\, .
\end{align}

\paragraph{NS5/D5 commutation relations.} Finally, let us work out the commutation relations between NS5s and D5s wrapping $\RP^4\subset \RP^5$. From the discussion above, we have that inserting a NS5 brane on $\Sigma^2\times\RP^4$ leads to a shift of $\diff H_3$ by $i(\PD_{\M{4}}[\Sigma^2]\smile \beta^{-1}(t_1))$. The $\diff F_7$ holonomy on the $\Xi^2\times\RP^4$ D5 worldvolume is invariant under this shift, but the D5 worldvolume theory contains couplings that do feel this shift: recall that the full Wess-Zumino couplings on the D5 brane on $\cM^6\subset X^{10}$ is the exponential of \cite{Cheung:1997az,Minasian:1997mm,Freed:2000ta}
\begin{align}
  \int_{\cM^6} e^{F-B} (C_0+C_2+C_4+C_6) \sqrt{\frac{\hat{A}(T\cM^6)}{\hat{A}(N\cM^6|_{X^{10}})}}\, .
\end{align}

We see that, crucially, there is a $B_2C_4$ contribution which will be
affected by the shift. The effect of the shift on this coupling is
best understood in the context of differential cohomology by
constructing a singular chain $\sC^7$ such that
$\partial \sC^7=\Xi^2\times \RP^4+\tilde \Xi^2\times \RP^4$, where
$\Xi^2$ and $\tilde\Xi^2$ are slightly displaced copies of the same
cycle (and have, in particular, the same orientation), and then
writing\footnote{Given than the integrand is valued in $\bT$
  multiplication by $\frac{1}{2}$ ultimately leads to dependence on a
  choice of quadratic refinement, see the comments below.}
\begin{align}
  \varphi = -\int_{\Xi^2\times \RP^4}B_2C_4 = -\frac{1}{2}\int_{\sC^7} \diff F_3\star \diff F_5\, .
\end{align}
Such a chain was constructed in section 4 of \cite{Witten:1998xy}, and from the discussion there it follows that
\begin{align}
  \int_{\sC^7} \diff H_3\star \diff F_5 =
  \left(\int_{\Xi^2\times\widetilde{\pt}}\diff H_3\right)
  \left(\int_{\RP^5}F_5\right)\, .
\end{align}
The last integral on the right is the number of mobile D3 branes on the orientifolded configuration, or in field theory terms the rank of the gauge group on the singularity.\footnote{It is only in the $\fso(2n)$ case that we have both NS5s and D5s as symmetry generators.} We will denote this integral by $n$. The action of $\diff H_3\to \diff H_3 + i(\PD_{\M{4}}[\Sigma^2]\smile \beta^{-1}(t_1))$ on $\varphi$ is therefore
\begin{align}
  \varphi\to\varphi - \frac{n}{2}\, \Xi^2\cdot \Sigma^2
  \int_{\widetilde{\pt}} \beta^{-1}(t_1) = \varphi - \frac{n}{4} \Xi^2\cdot\Sigma^2\, .
\end{align}
From here, we conclude that
\begin{align}
  \label{eq.NS5-D5-commutator}
  \text{NS5}(\Sigma^2)^{-1}\text{D5}(\Xi^2)\text{NS5}(\Sigma^2) = \exp\left(-2\pi i \frac{n}{4}\,\Xi^2\cdot\Sigma^2\right) \text{D5}(\Xi^2)\, .
\end{align}

\medskip

We note that in writing this formula for odd $n$ we have chosen a specific quadratic refinement of the holonomy term
\begin{align}
  \label{eq.refinement}
  \frac{1}{2} \int_{\widetilde{\pt}}\beta^{-1}(t_1) = \frac{1}{4} \mod 1\, .
\end{align}
(The other option would have been to choose $-1/4$ mod 1.) This choice
is meaningful whenever $n\notin 2\bZ$. The actual choice does not
affect the classification of global forms, but it appears to be
related to the structure of $SL(2,\mathbb{Z})$ duality orbits, see the
discussion in \S\ref{sec.k=2}. The fact that the definition of the
Chern-Simons terms in string theory requires a choice of quadratic
refinement is well known, see
\cite{Witten:1996md,Witten:1996hc,Diaconescu:2003bm,Freed:2004yc,Hopkins:2002rd,Hsieh:2020jpj}
for a sampling of detailed discussions, and we expect that the choice
of sign in~\eqref{eq.refinement} should follow from there.

\paragraph{D3/D3 commutation relations.} For completeness, let us mention the case of D3 branes wrapping $\RP^3\subset \RP^5$ and $\RP^1\subset\RP^5$. They lead to line and 3-surface excitations on the theory on $\M{5}$, and line and 3-surface operators acting on the boundary states. These branes are associated with 0-form and 2-form symmetries on the field theory. We can analyse this system via techniques very similar to the ones above, with the added simplification that we are now working in the untwisted sector. We find that
\begin{align}
  \text{D3}(\gamma\times\RP^3)^{-1}\text{D3}(\Sigma^3\times\RP^1)\text{D3}(\gamma\times\RP^3)
  = \exp\left(\pi i
    \gamma\cdot\Sigma^3\right)\text{D3}(\Sigma^3\times\RP^1)\, .
\end{align}

\subsection{An effective action on \alt{$\AdS_5$}{AdS5}}
\label{sec.Sfolds-effectiveaction}

We will now verify that the brane commutation relations that we have
obtained agree with those that can be derived from the action given
in~\cite{Bergman:2022otk}. We will discuss the $\fso(2n)$ case first,
and then discuss the modifications needed in the other cases. The
action in the $\fso(2n)$ case given in \cite{Bergman:2022otk} is (the
choice of sign is for convenience)
\begin{align}
  S_{\bfB}^{\fso(2n)} = -2 \pi i \int_{\M{5}} \bigl[ n
  \bfB_{\text{F1}} \wedge d \bfB_{\text{D1}}+ 2 \bfB_{\text{F1}} \wedge d
  \bfB_{\text{NS5}} + 2 \bfB_{\text{D1}} \wedge d \bfB_{\text{D5}} \bigr]\, .
  \label{eq.so(2N)-action}
\end{align}
We normalize the $U(1)$ $\bfB_I$ fields to have period 1 (as opposed
to the perhaps more standard convention in physics of period $2\pi$),
and $I$ stands for the type of brane which couples electrically to
$B_I$. So, for instance, $\bfB_{\text{D5}}$ should arise from
reduction of the $C_6$ RR field in supergravity. We can derive the
brane (non-)commutation relations from here following
\cite{Witten:1998wy}. We work in the path integral formulation (on a
euclidean spacetime), instead of the Hamiltonian formulation on a
constant time slice $\M{4}$ that we have been using so far. Assume
that we have two operators $U(\Sigma^2)$ and $V(\Xi^2)$, defined as in
section~\ref{sec.Sfolds-N=4-brane-noncommutativity} (we assume that
there are no torsional cycles on $\M{4}$, so we have chosen to replace
$\sigma_2$ by the Poincar\'e dual 2-cycle $\Sigma^2$ on $\M{4}$) such
that on $\M{4}$
\begin{align}
  U(\Sigma^2)^{-1} V(\Xi^2)U(\Sigma^2) = e^{-2\pi i\, q\,\Sigma^2\cdot
    \Xi^2} V(\Sigma^2)\, .
\end{align}
We interpret this as $V$ having charge $q$ under the abelian symmetry
generated by $U$, or equivalently as $V$ having charge $-q$ under $U$
(assuming that both operators are topological). The path integral
version of this statement is
\begin{align}
  U(\Sigma^2) V(\Xi^2) = e^{2\pi i \, q \, \sL_G(\Sigma^2, \Xi^2)} V(\Xi^2)
\end{align}
with $\sL_G(\Sigma^2, \Xi^2)$ the Gauss linking pairing on
$\M{5}$. This is the kind of relation that we aim to prove now,
starting from the action~\eqref{eq.so(2N)-action}.

\medskip

The presence of a brane of type $I$ wrapped on a 2-cycle
$C_I=\partial D_I$ leads to a factor
\begin{align}
  \exp \biggl( 2\pi i \int_{C_I} \bfB_I \biggr) = \exp\biggl(2\pi i \int_{D_I}
    d\bfB_I\biggr)
\end{align}
in the path integral. The equations of motion that follow from~\eqref{eq.so(2N)-action} in the presence of brane insertions are
\begin{align}
  \begin{aligned}
    - n d \bfB_{\text{D1}} - 2 d \bfB_{\text{NS5}} + \delta (C_\text{F1}) &= 0\,,\\
    n d \bfB_{\text{F1}} - 2 d \bfB_{\text{D5}} + \delta (C_\text{D1}) &= 0\,,\\
    2 d \bfB_{\text{F1}} + \delta (C_\text{NS5}) &= 0\,,\\
    2 d \bfB_{\text{D1}} + \delta (C_\text{D5}) &= 0\, ,
  \end{aligned}\label{eq.EOMeven}
\end{align}
where $C_{\text{F1}}$ is the cycle in $\M{4}$ wrapped by the F1 (which we can assume to be trivial in $\M{5}$ \cite{Witten:1998xy}), and similarly in the other cases. It is convenient to rewrite
\eqref{eq.EOMeven} as
\begin{align}
\begin{aligned}
d \bfB_{\text{NS5}} &= \frac{n}{4} \delta (C_\text{D5}) + \frac{1}{2} \delta (C_\text{F1}),\\
d \bfB_{\text{D5}} &= - \frac{n}{4} \delta (C_\text{NS5}) + \frac{1}{2} \delta (C_\text{D1}),\\
d \bfB_{\text{F1}} &= - \frac{1}{2} \delta (C_\text{NS5}),\\
d \bfB_{\text{D1}} &= - \frac{1}{2} \delta (C_\text{D5}) .
\end{aligned}
\label{eq.EOMevenconvenient}
\end{align}
From here we can easily compute the charges of one type of operator
under another. Consider for instance a D5 wrapping a curve
$C_{\text{D5}}$ that links
$\sL_G(C_{\text{D5}},C_{\text{NS5}})=D_{\text{NS5}}\cdot
C_{\text{D5}}$ times a curve $C_{\text{NS5}}$ wrapped by an NS5
branes. Assuming that there are no other branes in the problem, using
the equations of motion we have
\begin{align}
  \exp\! \biggl[ 2\pi i \int_{C_\text{NS5}} \!\!\!\! \bfB_{\text{NS5}} \biggr] \exp\! \biggl[ 2\pi i \int_{C_\text{D5}} \!\!\!\! \bfB_{\text{D5}} \biggr] & = \exp\! \biggl[ 2\pi i \int_{D_\text{NS5}} \!\!\!\! d\bfB_{\text{NS5}} \biggr] \exp\! \biggl[ 2\pi i \int_{C_\text{D5}} \!\!\!\! \bfB_{\text{D5}} \biggr] \nonumber \\
  & = \exp\! \biggl[ 2\pi i \frac{n}{4} D_{\text{NS5}}\cdot C_{\text{D5}} \biggr] \exp\! \biggl[ 2\pi i \int_{C_\text{D5}} \!\!\!\! \bfB_{\text{D5}} \biggr]
\end{align}
which indeed reproduces~\eqref{eq.NS5-D5-commutator}. The rest of the commutation relations give above can be derived similarly.

\subsection{Review of discrete torsion and Freed-Witten anomaly for \alt{$k=2$}{k=2}}
\label{sec:k=2-FW-review}

As discussed in \S\ref{sec.Sfolds-warmup} we may wrap 3 and 5 branes
on various cycles of the internal $\mathbb{RP}^5$, but in addition to
the topological restrictions there are quantum restrictions governed
by a type of Freed-Witten anomaly \cite{Freed:1999vc}, which were
worked out in \cite{Witten:1998xy}. In this section we review the
consequences of this anomaly for the symmetry structure of the
theory. We will discuss the $k=2$ case explicitly, a very similar
analysis holds for the $k>2$ cases discussed in the next section.

The NS and RR discrete torsions, $\theta_\text{NS}$ and
$\theta_\text{RR}$ take values of $0$ and $1/2$. There are thus four
choices of discrete torsion. A D5-brane can wrap an
$\mathbb{RP}^4$-cycle only if $\theta_\text{NS}=0$, and similarly an
NS5-brane can wrap a $\mathbb{RP}^4$-cycle only if
$\theta_\text{RR}=0$. Meanwhile, a D3-brane (without a string ending
on it) can wrap an $\mathbb{RP}^3$-cycle only if
$\theta_\text{RR}=\theta_\text{NS}=0$. A D3-brane can wrap an
$\mathbb{RP}^3$-cycle with a F1-string ending on it if and only if
$\theta_\text{RR}=1/2$ and $\theta_\text{NS}=0$. A 3-brane can wrap an
$\mathbb{RP}^3$-cycle with a D1 string ending on it if and only if
$\theta_{RR}=0$ and $\theta_\text{NS}=0$.

As argued in \cite{Witten:1998xy}, and reviewed in \S\ref{sec.k=2},
the algebras of the gauge theories correspond to the discrete torsion
in the following ways.
\begin{subequations}
	\begin{align}
	\fso(2N)&:\quad(\theta_\text{NS},\theta_\text{RR})=(0,0)\, ,\\
	\fso(2N+1)&:\quad(\theta_\text{NS},\theta_\text{RR})=(0,1/2)\, ,\\
	\fsp(2N)&:\quad(\theta_\text{NS},\theta_\text{RR})=(1/2,0)\, ,\\
	\fsp(2N)&:\quad(\theta_\text{NS},\theta_\text{RR})=(1/2,1/2)\, .
\end{align}
\end{subequations}

Let us now analyze in detail what happens to the symmetry generators
of the bulk theory in the presence of the discrete torsion. For
concreteness we focus on the case where
$(\theta_{\text{NS}},\theta_{\text{RR}})=(0,1/2)$. The other cases
follow analogously.

For this case, consider an F1 pushed to the boundary, which leads to
an insertion of a topological 2-surface in the field theory. The F1
string can end on a D3 brane wrapped on $\RP^3$, so that the
Freed-Witten anomaly on the D3 cancels by an explicit source
term. This implies that in the field field theory the topological
2-surface can end on a 1-surface. Since the topological 2-surface can
``open up'', it cannot measure any conserved 1-form symmetries, as any
line whose charge we would like to potentially measure can leak
through the holes we can nucleate on the 2-surface.

Suppose instead that the F1 ends on the boundary, by which we mean
that asymptotically it looks like $\bR\times\gamma$, with $\bR$ the
asymptotic radial direction in $\M{5}\sim \bR\times\M{4}$ and $\gamma$
a 1-chain in $\M{4}$. From the field theory perspective, this is a
Wilson line insertion on $\gamma$. The F1 can still end on a D3 brane,
which now implies that the Wilson line can end on a point
operator. Therefore, there is no operator under which the Wilson line
can carry charge. This has a simple origin in the bulk theory: due to
the Freed-Witten anomaly the NS5 brane cannot wrap a $\RP^4$-cycle for
this choice of torsional flux, and thus in this case there do not
exist wrapped NS5 branes to be the symmetry operators under which the
F1 string would be charged.

Finally we note that, while the boundary of the F1 along the wrapped
D3 cancels the Freed-Witten anomaly on the D3, we cannot use this
mechanism to cancel the Freed-Witten anomaly on a NS5. This follows
from Poincaré duality: the insertion in the internal space must be
along a Poincaré dual to the class induced by the flux, and on the NS5
on $\RP^4$ this is not a point (which is what the F1 generator wraps
in the internal $\RP^5$), but a line.

\subsection{Fluxes and Freed-Witten anomalies}
\label{sec.Sfolds-FW}

The analysis in the previous section can be generalised to the general
$\cN\geq 3$ S-fold case as follows. The background 3-form fluxes are
given by an element $\cF\in H^3(S^5/\bZ_k;
(\bZ\oplus\bZ)_{\rho_k})$. We note that this element is in all cases
torsional, and admits a flat differential cohomology uplift
$\diff \cF$. Let us assume that we now have a 5-brane wrapping a
representative $\Sigma$ of a class in the twisted homology
$H_4(S^5/\bZ_k;(\bZ\oplus\bZ)_{\rho_k})$. We represent this class by
its dual cohomology class
$\sigma\in H^1(S^5/\bZ_k; (\bZ\oplus\bZ)_{\rho_k})$, which can again
be uplifted to a flat differential cohomology class $\diff \sigma$. A
necessary condition for the Freed-Witten anomaly to vanish is that
\[
  \label{eq.Freed-Witten-linking}
  \sL(\diff \cF, \diff \sigma \star \diff \lambda) = 0 \mod 1
\]
for all $\lambda\in H^2(S^5/\bZ_k;\bZ)=\bZ_k$. The intuitive idea
behind this condition is that by probing by all $\lambda$ we can in
favourable circumstances detect whether the restriction of $\cF$ to
$\Sigma$ is trivial or not. If the restriction is trivial, the linking
pairing above will necessarily always vanish, and therefore if we
detect any non-trivial linking pairing we will have a Freed-Witten
anomaly. (The geometries wrapped by our branes are simple enough
that~\eqref{eq.Freed-Witten-linking} is in fact sufficient for
detecting absence of anomalies.)

We note that in general it is possible to have a term on the right
hand side of~\eqref{eq.Freed-Witten-linking} which does not depend on
$\diff\cF$, and only depends on the structure of the cycle wrapped by
the brane. For instance, it was shown in \cite{Freed:1999vc} that in
the case of a trivial $SL(2,\bZ)$ bundle over a D-brane wrapping a
cycle $\cM$ the right hand side is $\beta(w_2(T\cM))$, with $\beta$
the Bockstein associated to the short exact sequence
$0\to\bZ\to\bZ\to\bZ_2\to 0$. The orientifolded case was considered in
\cite{Witten:1998xy} (see also \cite{Hsieh:2020jpj}) where it was
conjectured that the right topological term on the right hand side was
$\tilde\beta(w_2(T\cM))$, with $\tilde\beta$ now associated to the
local version of the short sequence above, namely
$0\to\tbZ\to\tbZ\to\tbZ_2\to 0$, and we use the fact that
$\tbZ_2=\bZ_2$, so we can promote $w_2(T\cM)$ to a class in
$H^2(\cM;\tbZ_2)$.

The case of interest to us is $\cM=\RP^4$, and $w_2(T\RP^4)=0$,
so~\eqref{eq.Freed-Witten-linking} does not need to be modified, at
least in the $k=2$ case. (Assuming that the conjecture
in~\cite{Witten:1998xy} is correct.) This is also needed in order for
the holographic results to match the field theory results.

We can phrase the previous results for $k=2$ in a
$SL(2,\bZ)$-covariant way. Consider the short exact sequence of local
coefficients
$0\to \tbZ\oplus\tbZ\to \tbZ\oplus\tbZ \to \widetilde{\sC}_k\to 0$,
with associated Bockstein $\beta_{k}$. In the $k=2$ case we can embed
$w_2(T\cM)$ into
$H^2(\cM;\widetilde{\sC}_2)=H^2(\cM;\bZ_2)\oplus H^2(\cM;\bZ_2)$
simply by taking two copies. Denote the corresponding element of
$H^2(\cM;\widetilde{\sC}_2)$ by $\ssw_2$. (Recall that in the case of
interest to us we have $\ssw_2=0$.) It is then not hard to see that
we reproduce the Freed-Witten anomalies found in
\cite{Witten:1998xy} for the $k=2$ case if we
replace~\eqref{eq.Freed-Witten-linking} by
\[
  \sL(\diff \cF - \diff\beta_{k}(\ssw_2), \diff \sigma \star \diff
  \lambda) = 0 \mod 1\, .
\]

It is much less clear to us what to do in the $k>2$ case, so in the
rest of the analysis we will assume that, as in the $k\leq 2$ cases,
the na\"ive condition~\eqref{eq.Freed-Witten-linking} on $\diff\cF$ is
the correct one for all $k$, and no additional geometric contribution
appears for the branes we study.\footnote{We should be able to prove
  or refute this conjecture by lifting the 5-branes to M-theory M5
  branes, and analysing the analogous correction there. The form of
  the correction in M-theory was conjectured in \cite{Witten:1999vg}
  (see also
  \cite{Diaconescu:2003bm,Belov:2006jd,Monnier:2013kna,Monnier:2013rpa}). The
  relevant computation seems to be rather involved, and we will not
  attempt it here.}  We will momentarily give evidence in support of
this assumption in the $k=3$ case by showing that it leads to results
consistent with \cite{Zafrir:2020epd}.

Under this assumption, we can work out easily the $k>2$
cases. Consider for instance the $k=3$ case. Here we have a flux
$\cF\in H^3(S^5/\bZ_k; (\bZ\oplus\bZ)_{\rho_3})=\bZ_3$ and 5-branes
wrapping representatives $\Sigma$ of
$H_4(S^5/\bZ_k; (\bZ\oplus\bZ)_{\rho_3})=\bZ_3$. The
condition~\eqref{eq.Freed-Witten-linking} then projects out the
5-brane generating the 1-form symmetry if and only if there is a
non-trivial 3-form flux in the internal space.

These results support the claims in \cite{Zafrir:2020epd}. Two of the
$\cN=1$ theories constructed in that paper were argued to flow to
SCFTs in the same $\cN=1$ conformal manifolds as certain $\cN=3$
theories: a $k=3$ $\cN=3$ theory of rank one with non-trivial internal
flux (corresponding to the $G(1,1,3)$ complex reflection group, in the
notation of \cite{Aharony:2016kai}), and a $k=3$ $\cN=3$ theory of
rank three with no internal flux (corresponding to
$G(3,3,3)$). According to our analysis\footnote{We are assuming that
  our analysis remains valid for arbitrarily small rank. We expect
  this to be the case: while we have focused on the holographic
  description in order to stay close to the literature on the $\cN=4$
  case, the analysis of the ``SymTFT reduction'' along the lines of
  \cite{Apruzzi:2021nmk,Heckman:2022muc,vanBeest:2022fss} proceeds
  along nearly identical lines, and applies to the arbitrary rank
  case.}  we expect the first theory to have trivial 1-form symmetry
group, and the second to admit a global form having a $\bZ_3$ 1-form
symmetry group. The $\cN=1$ theories in proposed in
\cite{Zafrir:2020epd} do indeed exhibit these 1-form symmetry
structure.\footnote{One might hope to also compare the 0-form symmetry
  sector, and more generally the full non-invertible symmetry sector
  we describe below, but unfortunately the theories in
  \cite{Zafrir:2020epd} are not expected to flow to $\cN=3$ SCFTs,
  only to $\cN=1$ SCFTs in the same conformal manifold as the $\cN=3$
  SCFTs. The marginal operators that interpolate between both fixed
  points are expected to break the relevant discrete 0-form
  symmetries. We thank Gabi Zafrir for explaining this last point to
  us.}

\medskip

Finally, let us comment briefly on the 0-form symmetries of the
models. This was in fact already done in \cite{Aharony:2016kai}, so we
just quote their results (which were obtained by a similar reasoning
to the one above): in the absence of flux the $\cN=3$ theory
associated to the $\bZ_k$ quotient of $S^5$ has a $\bZ_k$ 2-form
symmetry, generated by D3 branes wrapping the $\bZ_k$ generators of
$H_3(S^5/\bZ_k; \bZ)$, for appropriate boundary conditions. Different
choices of boundary conditions lead to 0-form symmetries, and the D3
branes just described end on the boundary, giving the point operators
charged under the 0-form symmetry. Non-trivial fluxes make these brane
wrappings suffer from a Freed-Witten anomaly, and the boundary SCFTs
do not have the 0-form symmetry any longer.

\subsection{'t Hooft anomalies and non-invertible symmetries}
\label{sec.Sfolds-mixedanomalies}

The action given in~\eqref{eq.so(2N)-action} does not include a term
involving the outer automorphism 0-form symmetry of the $\Spin(2N)$
theory. To account for this we add two $U(1)$ fields $\bfA_1$ and
$\bfA_3$ with a coupling
\[
  S^{\fso(2n)}_{\bfA} = 4\pi i \int_{\M{5}} \bfA_1 \wedge d\bfA_3\, . \label{eqn:A1A3action}
\]
This theory can also be described as a $\bZ_2$ gauge theory with action
\[
  S^{\fso(2n)}_{A} = \pi i \int_{\M{5}} A_1\smile \delta A_3
\]
where the $A_1$, $A_3$ fields are $\bZ_2\df \bZ/2\bZ$ fields related
to the continuous fields by $A_i\df 2\bfA_i$ (reducing coefficients
appropriately from $\bR/2\bZ$ to $\bZ_2$).

The operators in this theory behave precisely as the D3 branes wrapped
on $\RP^1$ and $\RP^3$ described above. More generally, for $k>2$ we
have \cite{Aharony:2016kai}
\[
  S^{k}_{\bfA} = 2\pi i k \int_{\M{5}} \bfA_1 \wedge d\bfA_3\, ,
\]
or in terms of finite fields $A_i\df k\bfA_i$
\[
  S^{k}_{\sA} = \frac{2\pi i}{k} \int_{\M{5}} A_1 \smile\delta A_3\, .
\]

This is nevertheless not the full answer: as discussed in
\cite{Hsin:2020nts} (see also \cite{Cordova:2017vab}), the interplay
of the outer automorphism 0-form symmetry with the 1-form symmetry
sector is subtle. For instance, in the $SO(2n)$ theory there is a
mixed 't Hooft anomaly involving 0-form and 1-form symmetries, which
leads to 2-group symmetries in the $\Spin(2n)$ case upon gauging the
magnetic 1-form symmetry \cite{Tachikawa:2017gyf}. Other choices of
gauging lead to non-invertibles symmetries, see
\cite{Bhardwaj:2022yxj}. In this section we would like to reproduce
some of these results from the holographic perspective, and extend
them to the $\cN=3$ case, where we also find non-invertible
symmetries.\footnote{This adds to the large amount of evidence that
  has accumulated during the last year showing that non-invertible
  symmetries appear in many interesting physical theories. We refer
  the reader to
  \cite{Frohlich:2009gb,Carqueville:2012dk,Brunner:2013xna,Bhardwaj:2017xup,Gaiotto:2019xmp,Heidenreich:2021xpr,Choi:2021kmx,Kaidi:2021xfk,
    Roumpedakis:2022aik,
    Bhardwaj:2022yxj,Arias-Tamargo:2022nlf,Choi:2022zal,Choi:2022jqy,
    Cordova:2022ieu,Kaidi:2022uux,Antinucci:2022eat,Bashmakov:2022jtl,Damia:2022rxw,Damia:2022bcd,Bhardwaj:2022lsg,Lin:2022xod,Bartsch:2022mpm,Apruzzi:2022rei,GarciaEtxebarria:2022vzq,Lu:2022ver,Heckman:2022muc,Niro:2022ctq,Kaidi:2022cpf,Mekareeya:2022spm,Antinucci:2022vyk,Giaccari:2022xgs,Bashmakov:2022uek,Cordova:2022fhg,Karasik:2022kkq,GarciaEtxebarria:2022jky,Choi:2022fgx,Bhardwaj:2022kot,Bhardwaj:2022maz,Bartsch:2022ytj,Das:2022fho,Heckman:2022xgu,Antinucci:2022cdi,Apte:2022xtu,Garcia-Valdecasas:2023mis,Kaidi:2023maf}
  for some of the recent developments in this field.} Our analysis is
incomplete in two important respects: first, we will restrict
ourselves to the case of $n=2k\in 2\bZ$, with vanishing discrete
$\theta$ angle. And we will focus on deriving the anomaly of the
$SO(4k)$ theory. Since other global forms are obtained by gauging, we
expect that the bulk dynamics is the same in all cases, but it is
still important to understand the structure of the field theory
symmetries directly from the holographic dual for all global
forms. The non-invertible symmetries for the theories at hand were
understood in \cite{GarciaEtxebarria:2022vzq} (see also
\cite{Apruzzi:2022rei,Heckman:2022muc} for closely related results),
and below we generalise this analysis to the $\cN=3$ case. To our
knowledge the case of 2-group symmetries has not yet been fully
understood in holographic terms, but see
\cite{DelZotto:2020sop,Apruzzi:2021phx,Apruzzi:2021vcu,Bhardwaj:2021wif,DelZotto:2022fnw,DelZotto:2022joo,Cvetic:2022imb}
for progress on understanding 2-groups in various string
constructions.

In general, extending slightly the discussion in
\cite{Apruzzi:2021nmk}, given a (possibly twisted) differential
cohomology cochain $\diff F$ on $\M{5}\times \RP^5$, up to
topologically trivial terms we can express its leading contribution to
the effective theory on $\M{5}$ in terms of torsional differential
cohomology classes on $\RP^5$ times differential cochains on $\M{5}$.
We have
\[
  \label{eq.H3-F3-expansion}
  \diff H_3 = \diff B_{\text{F1}}\star \diff t_1^{\text{NSNS}} + \diff
  \theta_{\text{NS}} \star \diff t_3^{\text{NSNS}} \qquad ; \qquad \diff
  F_3 = \diff B_{\text{D1}}\star \diff t_1^{\text{RR}} + \diff
  \theta_{\text{RR}}\star \diff t_3^{\text{RR}}
\]
in the twisted sector, with
$\diff t_i^{\text{NSNS}},\diff t_i^{\text{RR}}$ flat differential
cohomology uplifts of the generators of
$H^i(\RP^5;(\tbZ\oplus\tbZ)_{\rho_2})$, and
\[
  \label{eq.F5-expansion}
  \diff F_5 = \diff A_1 \star \diff u_4 + \diff A_3 \star \diff u_2 + \ldots
\]
in the untwisted sector ($\diff u_4$ is a flat uplift of the generator
of $H^i(\RP^5;\bZ)$). We have omitted some terms in $\diff F_5$
proportional to $n$ that do not enter in the computation of the
anomaly.

The IIB pseudo-action on $\M{5}\times \RP^5$ includes a term of the form
\[
  \label{eq.F5-H3-F3}
  S_{\text{IIB}} = 2\pi i \int_{\M{5}\times \RP^5} \diff F_5\star
  \diff H_3 \star \diff F_3\, .
\]
Introducing the expansions above into this expression, and using
\[
  \int_{\RP^5} \diff u_4 \star \diff t_1^{\text{RR}} \star \diff
  t_1^{\text{NSNS}} = \frac{1}{2} \mod 1
\]
(which we can derive using similar arguments to the ones given above,
see the general derivation below) we obtain the effective coupling
\[
  \label{eq.so(4k)-anomaly}
  S_{\text{anomaly}} = \pi i \int_{\M{5}} A_1\smile B_{\text{F1}} \smile
  B_{\text{D1}} \, .
\]
Here we have used that this integral is a primary invariant to express
it in terms of more conventional cochain integrals.  If we choose
Dirichlet boundary conditions for $A_1$, $B_{\text{F1}}$ and
$B_{\text{D1}}$, which corresponds to the $SO(2n)$ choice of global
form, this is precisely the anomaly theory described
in~\cite{Hsin:2020nts}.

\medskip

The coupling~\eqref{eq.F5-H3-F3} can be written more covariantly as
\[
  \label{eq.F-F^2 covariant}
  S_{\text{IIB}} = -\frac{2\pi i}{2} \int_{\M{5}\times (S^5/\bZ_k)} \diff
  F_5 \star \diff \cF \star \diff \cF\, .
\]
where the factor of $-1/2$ encodes the fact that we are dealing with a
quadratic refinement, as we will explain momentarily. The overall
choice of sign is conventional, we have chosen the sign that agrees
with existing conventions in M-theory, where the dual coupling arises
from expanding $-\frac{1}{6}\int \diff G_4^3$, with
$\diff G_4 = \diff F_4 + \diff \cF_4$. Here $\diff F_4$ and
$\diff \cF_4$ are the M-theory duals of $F_5$ and $\diff F$
respectively. The subtleties in dealing with this fractional prefactor
have been extensively discussed in the M-theory setting, starting with
\cite{Witten:1996hc,Witten:1996md}.

For $k>2$, switching on a background for the 1-form symmetry
corresponds to taking
\[
  \diff \cF = \diff B_2 \star \diff t_1
\]
with $\diff t_1$ a flat uplift of the single generator of
$H^1(S^5/\bZ_k;(\bZ\oplus\bZ)_{\rho_k})=\sC_k$. Expanding $\diff F_5$
as in~\eqref{eq.F5-expansion}, and integrating, we find an effective
coupling
\[
  \label{eq:S-fold-anomaly}
  S_{\text{anomaly}} = 2\pi i \ssq_k \int_{\M{5}} A_1\smile B_2 \smile
  B_2 \, ,
\]
with
\[
  \ssq_k = -\frac{1}{2}\int_{S^5/\bZ_k} \diff u_4 \star \diff t_1\star
  \diff t_1 = \begin{cases}
    -\frac{1}{3} & \text{for } k=3\, ,\\
    -\frac{1}{4} & \text{for } k=4\, ,\\
    \phantom{-}0 & \text{for } k=6\, ,
  \end{cases}
\]
and the anomaly theory~\eqref{eq.so(4k)-anomaly} for $k=2$. (With a
caveat to be discussed below.) We note that the sign of $\ssq_k$ can
be absorbed in a redefinition of $A_1$, or in the F-theory formulation
in a sign redefinition of $\diff C_3$.

To compute $\ssq_k$, consider the F-theory description of the system,
where we have
\[
  \ssq_k = -\frac{1}{2}\int_{X^7_k} \diff u_4 \star \diff t_2 \star
  \diff t_2\, .
\]
Here $X^7_k$ is an elliptically fibered 7-fold over $S^5/\bZ_k$
encoding the data of the $SL(2,\bZ)$ fibration over $S^5/\bZ_k$. We
have abused notation slightly and also called $u_4$ the pullback to
$H^4(X^7_k;\bZ)$ of the generator of $H^4(S^5/\bZ_k;\bZ)$, which in
turn is the pullback of the generator of $H^4(\bP^2;\bZ)$. (This last
statement follows easily from the Gysin exact sequence.) The class
$t_2\in H^2(X^7_k;\bZ)$ should be understood as the uplift to M-theory
of $t_1\in H^1(S^5/\bZ_k;(\bZ\oplus\bZ)_{\rho_k})$. Mathematically,
what we mean by ``uplift'' is that the given element of
$H^1(S^5/\bZ_k;(\bZ\oplus\bZ)_{\rho_k})$ survives to a non-trivial
element of $H^2(X^7_k;\bZ)$ in the Leray-Serre spectral sequence
(keeping in mind $H^1(T^2;\bZ)=(\bZ\oplus\bZ)_{\rho_k}$ as a local
coefficient system).

A useful alternative viewpoint on $X^7_k$ is that it is a fibration of
$\sM_k$ (defined in \S\ref{sec.Sfolds-kgeneral} above) over
$\bP^2$. The class $u_4$ is still a pullback of the fundamental class
of $\bP^2$ to the total space of the fibration, while $t_2$ arises
from the generator of $H^2(\sM_k;\bZ)=\sC_k$ in the Leray-Serre
spectral sequence. By a push-pull argument, this implies that $\ssq_k$
reduces to a quadratic refinement of the Chern-Simons coupling on
$\sM_k$, which we can write (again abusing notation slightly, by
denoting $t_2$ the generator of $H^2(\sM_k;\bZ)$ and $\diff t_2$ a
flat uplift to differential cohomology)
\[
  \ssq_k = -\frac{1}{2}\int_{\sM_k} \diff t_2\star \diff t_2\, .
\]
This kind of quadratic refinement, in a situation very analogous to
ours, was computed in~\cite{Apruzzi:2021nmk}. The idea is to construct
a (Calabi-Yau) manifold $K^4_k$ such that $\partial K^4_k=\sM_k$, and
then translate the computation of the quadratic refinement of the
Chern-Simons term to a problem in intersection theory on $K_k^4$. We
refer the reader to \cite{Apruzzi:2021nmk} for details. In our case
there is a natural choice of $K_k^4$: consider an elliptic fibration
over $\bC$ with a Kodaira singularity at $z=0$ of type $IV^*$ (leading
to an $E_6$ gauge theory, if this was an M-theory background). We take
$K^4_3$ to be the total space of the fibration over a disk
$D=\{|z|\leq 1\}$. Recalling that the monodromy around the singularity
is precisely $\rho_3$, we have $\partial K^4_3 = \sM_k$. Similarly,
$K_4^4$ comes from an $E_7$ singularity, $K_6^4$ from an $E_8$
singularity, and $K_2^4$ from a $D_4$ singularity. Repeating the
computation of \cite{Apruzzi:2021nmk} for these geometries leads to
the results for $\ssq_k$ stated above. (The local form of the geometry
close to the singular locus is of the form $\bC^2/\Gamma$, with
$\Gamma\subset SU(2)$, which are some of the cases studied in that
paper, so the details are essentially identical.)

The $k=2$ case requires some additional comments. In this case we have
$H^2(\sM_k;(\bZ\oplus\bZ)_{\rho_k})=\bZ_2\oplus\bZ_2$. We denote the
two generators $t_2^{\text{NSNS}}$ and $t_2^{\text{RR}}$. From the
arguments above one obtains
\[
  \label{eq.k=2-quadratic-refinement}
  \ssq_2 = -\frac{1}{2} \int_{\sM_2} \diff t_2^{\text{NSNS}}\star \diff t_2^{\text{NSNS}} = -\frac{1}{2} \int_{\sM_2} \diff t_2^{\text{RR}}\star \diff t_2^{\text{RR}} = \frac{1}{2} \mod 1
\]
and
\[
  -\frac{1}{2} \int_{\sM_2} \diff t_2^{\text{NSNS}} \star \diff t_2^{\text{RR}} + \diff t_2^{\text{RR}} \star \diff t_2^{\text{NSNS}} = \frac{1}{2} \mod 1\, .
\]
The second integral leads to~\eqref{eq.so(4k)-anomaly} when we expand
as in~\eqref{eq.H3-F3-expansion} (replacing $t_1$ by $t_2$)
and~\eqref{eq.F5-expansion}, as advertised. On the other hand,
\eqref{eq.k=2-quadratic-refinement} would lead to contributions to the
anomaly theory of the form
\[
  S_{\text{anomaly}} = \pi i \int_{\M{5}} \bigl[ A_1\smile B_{\text{F1}} \smile B_{\text{F1}} +  A_1\smile B_{\text{D1}} \smile B_{\text{D1}} \bigr]\, .
\]
These couplings in fact vanish under the assumption that $\M{5}$ is
Spin and has no torsion. First, note that since $\M{5}$ has no
torsion, $B_{\text{F1}}$ is necessarily the mod 2 reduction of a class
in $H^2(\M{5};\bZ)$. This implies, in particular, that
$\Sq1(B_{\text{F1}})=\rho_2(\beta(B_{\text{F1}}))=0$, where $\rho_2$
indicates reduction modulo 2 and $\beta$ is the Bockstein homomorphism
associated to $0\to\bZ\to\bZ\to\bZ_2\to 0$. We therefore have (viewing
all classes as living in singular cohomology with $\bZ_2$
coefficients)
\begin{align}
  \int_{\M{5}} A_1 \smile B_{\text{F1}} \smile B_{\text{F1}} & = \int_{\M{5}}
  A_1 \smile \Sq2(B_{\text{F1}}) = \int_{\M{5}} \Sq2(A_1\smile B_{\text{F1}}) \nonumber \\
  & = \int_{\M{5}} \nu_2 \smile A_1\smile B_{\text{F1}} = 0\, ,
\end{align}
with $\nu_2$ the second Wu class of $\M{5}$, which vanishes since $\M{5}$
is a Spin manifold by assumption. It would be interesting to
understand what is the fate of these couplings when we relax our
assumptions on $\M{5}$.

\subsubsection*{Non-invertible symmetries} \label{sec.Sfolds-non-invertibility}

The existence of the cubic anomalies~\eqref{eq.so(4k)-anomaly}
and~\eqref{eq:S-fold-anomaly} implies the existence of non-invertible
symmetries upon gauging the 1-form symmetries appearing in the
anomaly, as argued in~\cite{Kaidi:2021xfk} and further studied in
\cite{Bhardwaj:2022yxj}. In this case the generator
\[
  U(\Sigma_3) = \exp\biggl(\pi i \int_{\Sigma_3} A_3\biggr)
\]
of the 0-form symmetry in the $SO(4k)$ theory is not invariant under
gauge transformations in the 1-form symmetry sector, so the operator
does not survive in the theory where the full $\bZ_2\times\bZ_2$
1-form symmetry of the $SO(4k)$ theory has been gauged (which by a
suitable choice of conventions we can call the $Sc(4k)$ theory). The
fundamental observation of \cite{Kaidi:2021xfk,Bhardwaj:2022yxj} is
that we can still construct a gauge invariant operator by stacking
$U(\Sigma_3)$ with an $\bZ_2$ gauge theory coupling to the 1-form
background in a suitably anomalous way:
\[
  \label{eq:non-invertible}
  \cN(\Sigma_3) = U(\Sigma_3)\exp\biggl[\pi i\int_{\Sigma_3} \bigl(\gamma_1\smile \delta\phi_1 + \gamma_1 \smile B_{\text{D1}} + \phi_1 \smile B_{\text{F1}}\bigr)\biggr]\,.
\]
Here $\gamma_1$ and $\phi_1$ are dynamical (topological) fields living
on the defect, which should be integrated over. The operator
$\cN(\Sigma_3)$ is now invariant under gauge transformations of the
1-form symmetry backgrounds $B_{\text{D1}}$ and $B_{\text{F1}}$, and
survives in the $Sc(4k)$ theory. The price to pay is that
$\cN(\Sigma)^{-1}$ does no longer exist (we say that $\cN(\Sigma_3)$
is \emph{non-invertible}), and in particular
$\cN(\Sigma_3)\times \cN(\Sigma_3)^{\dagger}$ is not the identity but
rather a condensation defect \cite{Roumpedakis:2022aik} (see also
\cite{Kong:2014qka,Gaiotto:2019xmp,kong2020algebraic}).

The analysis in \cite{Kaidi:2021xfk} extends straightforwardly to the
$\cN=3$ S-folds that we studied above: due to the cubic
anomaly~\eqref{eq:S-fold-anomaly} the 0-form symmetry generator will
not survive in the theory where we gauge the 1-form symmetry, but if a
suitably anomalous dressing exists then this might lead to a
non-invertible symmetry generator.

Interestingly, a number of recent works
\cite{Apruzzi:2022rei,GarciaEtxebarria:2022vzq,Heckman:2022muc} have
argued that the anomalous $BF$ theory in~\eqref{eq:non-invertible}
arises from reducing the Chern-Simons action on the symmetry
generating D-branes along torsional cycles in the internal space. It
is natural to guess that this is also the case for the cases studied
here. Luckily, the relevant analysis has already been done in
\cite{GarciaEtxebarria:2022vzq} for the $k=2$ case and in
\cite{Heckman:2022muc} for the $k>2$ case. (The theories studied in
\cite{Heckman:2022muc} were non-Higgsable clusters in $d=6$ and not
$\cN=3$ S-folds, but the relevant computation is identical in both
cases.) In all cases one can see that the resulting theory is
anomalous in precisely the right way to lead to a gauge-invariant
defect in the gauged theory. For instance, it follows from the
discussion in \cite{Heckman:2022muc} that in the $k>2$ cases, after
reduction on the torsional cycle in the internal space, there is an
effective Chern-Simons theory on the dynamical symmetry generator with
action\footnote{We are very thankful to J.~Heckman for pointing out
  that the results in \cite{Heckman:2022muc} would be very useful for
  our analysis, and for discussions on the normalisation of the
  $c$-dependent terms in~\eqref{eq:D3-S-fold-CS}.}
\[
  \label{eq:D3-S-fold-CS}
  S_{\text{CS}} = 2\pi i \int \biggl[\frac{1}{k}A_3 + \fq_k c\smile \delta c + 2 \fq_k B_2 \smile c \biggr]\,,
\]
with $c$ a dynamical cocycle valued on $\sC_k$, and $A_3$, $B_2$
background fields as above. (The difference in the normalisation of
the $A_3$ dependent term with respect to \cite{Heckman:2022muc} is due
to the difference in the setups we are considering.) The anomalous
variation of this action under gauge transformations for $B_2$
precisely cancels the anomalous variation due
to~\eqref{eq:S-fold-anomaly}.

\subsubsection*{Freed-Witten anomalies as a Stückelberg mechanism}

Finally, let us briefly comment on the effective action in the case
with 3-form flux in the internal space, namely, with
$\theta_{\text{RR}}$ or $\theta_{\text{NS}}$ different from 0. We
emphasise that the discussion in this section is not needed for the
rest of the paper, but it provides a nice alternative viewpoint on the
operator-centered viewpoint that we have adopted in most of the paper.

From the expansion~\eqref{eq.H3-F3-expansion} and
\[
  \int_{\RP^5} \diff t_1 \star \diff t_3 \star \diff u_2 = \frac{1}{2} \pmod 1 \,,
\]
integration of~\eqref{eq.F5-H3-F3} over $\RP^5$ leads to couplings of
the form\footnote{We are ignoring the mixed 't Hooft anomaly
  contribution here, which leads to subtleties in the definition of
  the Stückelberg action that we do not fully understand.}
\[
  S_{\theta} = \pi i \int_{\M{5}} \bigl[A_3 \smile \theta_{\text{NS}} \smile
  B_{\text{D1}} + A_3 \smile \theta_{\text{RR}} \smile B_{\text{F1}}\bigr]\, .
\]
In continuous notation, this modifies the combined
action~\eqref{eq.so(2N)-action} and~\eqref{eqn:A1A3action} to
\begin{multline}
  S_{\bfB,\theta}^{\fso(2n)} = -2 \pi i \int_{\M{5}} \bigl[ n
  \bfB_{\text{F1}} \wedge d \bfB_{\text{D1}}+ 2 \bfB_{\text{F1}} \wedge
  (d \bfB_{\text{NS5}} - \theta_{\text{RR}}\bfA_3) \\ + 2 \bfB_{\text{D1}} \wedge (d \bfB_{\text{D5}}- \theta_{\text{NS}}\bfA_3) +2 \bfA_1 \wedge d \bfA_3 \bigr] \, . \label{eqn:Stuckelbergaction}
\end{multline}
The presence of the Stückelberg terms gives a nice effective field
theory reinterpretation of the fact, pointed out above, that some of
the symmetry operators are absent in the presence of background 3-form
fluxes.

\section{The \alt{$\cN=4$}{N=4} theories from the \alt{$k=2$}{k=2} S-fold}\label{sec.k=2}

Specializing to $k=2$, we now compare the results of the above analysis with existing results on $\fso$ and $\fsp$
$\cN=4$ gauge theories
\cite{Witten:1998wy,Witten:1998xy,Aharony:2013hda,GarciaEtxebarria:2022vzq,Bergman:2022otk}, yielding an important cross check of our methods.

To do so, we start with the dictionary between branes and line operators established in
\cite{Witten:1998wy,Witten:1998xy,Bergman:2022otk}. Applying this dictionary to the bulk-commutation relations derived in \S\ref{sec.Sfolds-N=4-brane-noncommutativity},  we reproduce the field-theory mutual locality relations of \cite{Aharony:2013hda},
thereby demonstrating that the allowed global structures of $\cN=4$ S-folds perfectly agrees with the known global structures of $\cN=4$ theories with $\fso$ and $\fsp$ gauge
algebras. We also discuss how the 1-form symmetries can be
understood 
using the bulk effective action considered in \S\ref{sec.Sfolds-effectiveaction}, \S\ref{sec.Sfolds-mixedanomalies}. As a
final check, we conclude by showing that the
$\SL{2,\bZ}$ orbits of the bulk theory agree with the field theory
orbits described in \cite{Aharony:2013hda}.

Let us fix the action of S and T generators of $\SL{2,\bZ}$ in string theory to be
\begin{align}
	\begin{aligned}
		\text{F1}&\xrightarrow{\text{S}}\overline{\text{D1}},\quad
		&\text{D1}&\xrightarrow{\text{S}}\text{F1},\quad
		&\text{NS5}&\xrightarrow{\text{S}}\overline{\text{D5}},\quad
		&\text{D5}&\xrightarrow{\text{S}}\text{NS5},\\
		\text{F1}&\xrightarrow{\text{T}}\text{F1},\quad
		&\text{D1}&\xrightarrow{\text{T}}\overline{\text{F1}}+\text{D1},\quad
		&\text{NS5}&\xrightarrow{\text{T}}\text{NS5}+\text{D5},\quad
		&\text{D5}&\xrightarrow{\text{T}}\text{D5},
	\end{aligned}\label{eq.SdualsBulk}
\end{align}
where the bar over a brane denotes an antibrane, and the sum of two branes is interpreted as a bound state. 

\subsection{The line operator dictionary}\label{sec.k=2-Dictionary}

The (non-topological) Wilson and 't Hooft lines of $\mathcal{N}=4$ gauge theories are described holographically as the boundaries of dynamical strings in the AdS$_5$ dual. The latter can arise either from ten-dimensional strings or from ten-dimensional five-branes wrapped on (torsion) four cycles. The dictionary between the two was first worked out for $\fsu$ theories (i.e., for $k=1$) in \cite{Witten:1998wy,Witten:1998xy}. Here we focus on the $k=2$ dictionary for $\fso$ and $\fsp$ theories obtained in \cite{Witten:1998wy,Witten:1998xy,Bergman:2022otk} and summarized in table \ref{table:dictionary}.

\begin{table}
\begin{align*}
\begin{array}{c|c| c| c|c|c}
&&\text{F1}&\text{D1}&\text{D5}&\text{NS5}\\
\hline
&\text{Rep}_e\otimes\text{Rep}_m &\text{Vect}\otimes1&1\otimes\text{Vect}&\text{Spin}\otimes1&1\otimes\text{Spin}\\
\hline
\fso(2n+1)&(z_e,z_m)\in (\mathbb Z_2\times \mathbb Z_2)&(0,0)&(0,1)&(1,0)&\text{absent}\\
\fsp(2n)_{\theta_{\text{RR}}=0}&(z_e,z_m)\in (\mathbb Z_2\times \mathbb Z_2)&(1,0)&(0,0)&\text{absent}&(0,1)\\
\fso(4k+2)&(z_e,z_m)\in (\mathbb Z_4\times \mathbb Z_4)&(2,0)&(0,2)&(1,0)&(0,\pm1)\\
\fso(4k)&\begin{matrix}(z_{e,S},z_{e,C};z_{m,S},z_{m,C})\\ \in(\mathbb Z_2\times \mathbb Z_2)\times(\mathbb Z_2\times \mathbb Z_2)\end{matrix}&(1,1;0,0)&(0,0;1,1)&(1,0;0,0)&(0,0;1,0)\\
\end{array}
\end{align*}
\captionof{table}{The dictionary between field-theory lines and their dual
  bulk-worldsheets from
  \cite{Witten:1998wy,Witten:1998xy,Bergman:2022otk}. Note that for $\fso(4k+2)$ the
  sign of the 1-form charge of the NS5 brane relative to that of the D5 brane is not fixed by our analysis,
  though we believe it is related to the choice of quadratic
  refinement, see the comments in
  \S\ref{sec.Sfolds-N=4-brane-noncommutativity}. Matching with
  \cite{Aharony:2013hda} requires this charge to be $(0,-1)$ for
   $\fso(8k+2)$ and $(0,1)$ for $\fso(8k+6)$.
\label{table:dictionary}}
\end{table}

Let us describe briefly how this dictionary is worked out. Since an F1 string ending on the
$\mathcal{M}^4$ boundary is dual to a Wilson line in the vector
representation of the gauge group, applying S-duality simultaneously to the bulk and boundary theories implies that a D1 string ending on
$\mathcal{M}^4$ is dual to an 't Hooft line in the vector
representation of the dual gauge group. Meanwhile, it was argued in
\cite{Witten:1998xy} that the boundary of a D5 brane wrapping a
$\RP^4$ cycle in $\RP^5$ and ending on $\mathcal{M}^4$ is dual to a
Wilson line in the spinor representation of the gauge group. (As
reviewed in \secref{sec:k=2-FW-review} this is only possible when the
gauge algebra is $\fso(n)$, i.e. when $\theta_{\text{NS}}=0$.)
Thus, by S-duality the boundary of an NS5 brane wrapping
a $\RP^4$ cycle in $\RP^5$ and ending on $\mathcal{M}^4$ is dual
to an 't Hooft line in the spinor representation of the Langlands dual
gauge group.

Turning on discrete torsion restricts which five-branes can wrap $\RP^4$, hence in the $\fso(2n+1)$ and $\fsp(2n)$ theories, only one of the D5,
NS5, and D5$+$NS5 branes will be present as a line operator in the dual
theory. Likewise, depending on the torsion either the F1, the D1 or the F1$+$D1 can end on a wrapped D3 brane, hence the corresponding line operator is no longer charged under a 1-form symmetry.

\subsection{Field theory mutual locality from bulk non-commutativity}\label{sec.k=2-MutualLocality}

We now use the dictionary above, together with the bulk brane
commutation relations derived in
\secref{sec.Sfolds-N=4-brane-noncommutativity}, to reproduce the mutual
locality relations obtained used field theory methods in
\cite{Aharony:2013hda}.

Label an arbitrary bound state of $n_\text{F1}$ F1 strings, $n_\text{D1}$ D1 strings, etc., by
\begin{equation}
[n_{\text{F1}},n_{\text{D1}},n_{\text{D5}},n_{\text{NS5}}]=(n_{\text{F1}}\text{F1}+n_{\text{D1}}\text{D1}+n_{\text{D5}}\text{D5}+n_{\text{NS5}}\text{NS5}).
\label{eq.boundnotation}
\end{equation}
In \secref{sec.Sfolds-N=4-brane-noncommutativity}, we obtained the following commutation relations in the $k=2$ S-fold
\begin{subequations}
\begin{align}
\text{F1}(\Sigma^2)\text{D1}(\Xi^2)&=\text{D1}(\Xi^2)\text{F1}(\Sigma^2)\,,\\
\text{F1}(\Sigma^2)\text{D5}(\Xi^2)&=\text{D5}(\Xi^2)\text{F1}(\Sigma^2)\,,\\
\text{F1}(\Sigma^2)\text{NS5}(\Xi^2)&=\exp\left(\frac{2\pi i}{2}\Sigma^2\cdot\Xi^2\right)\text{NS5}(\Xi^2)\text{F1}(\Sigma^2)\,,\\
\text{D1}(\Sigma^2)\text{D5}(\Xi^2)&=\exp\left(\frac{2\pi i}{2}\Sigma^2\cdot\Xi^2\right)\text{D5}(\Xi^2)\text{D1}(\Sigma^2)\,,\\
\text{D1}(\Sigma^2)\text{NS5}(\Xi^2)&=\text{NS5}(\Xi^2)\text{D1}(\Sigma^2)\,,\\
\text{D5}(\Sigma^2)\text{NS5}(\Xi^2)&=\exp\left(\frac{2\pi i n}{4}\Sigma^2\cdot\Xi^2\right)\text{NS5}(\Xi^2)\text{D5}(\Sigma^2)\,.
\end{align}
\end{subequations}
Since each commutation produces at most a phase, the same is true for an arbitrary bound state, for which we obtain
\begin{multline}
	\bigl([m_{\text{F1}},m_{\text{D1}},m_{\text{D5}},m_{\text{NS5}}](\Xi^2)\bigr)^{-1}[n_{\text{F1}},n_{\text{D1}},n_{\text{D5}},n_{\text{NS5}}](\Sigma^2)\,[m_{\text{F1}},m_{\text{D1}},m_{\text{D5}},m_{\text{NS5}}](\Xi^2)\\=\exp\biggl[2\pi i\biggl(\frac{n_{\text{F1}}m_{\text{NS5}}-n_{\text{NS5}}m_{\text{F1}}}{2}+\frac{n_{\text{D1}}m_{\text{D5}}-n_{\text{D5}}m_{\text{D1}}}{2}+\frac{(n_{\text{D5}}m_{\text{NS5}}-n_{\text{NS5}}m_{\text{D5}})n}{4}\biggr)\biggr]\\ \cdot
[n_{\text{F1}},n_{\text{D1}},n_{\text{D5}},n_{\text{NS5}}](\Sigma^2).\label{eq.boundcommutators}
\end{multline}

The bulk theories are selected by a choice of boundary condition on the fields. Commutation relations of the fields constrain the possible boundary conditions \cite{GarciaEtxebarria:2019caf}, and consistent choices thereof are equivalent to mutually commuting sets of operators.

\subsubsection*{Deriving mutual locality conditions for the $\fso(4j)$ case} \label{sec.Sfolds-fso(4j)}
We now demonstrate how to derive the mutual locality conditions in field-theory from the bound-state commutators \eqref{eq.boundcommutators}. We focus here on the $\fso(4j)$ case with $\theta_{\text{NS}}=0=\theta_{\text{RR}}$. The $\fso(4j+2)$, $\fsp(2n)$, and $\fso(2n+1)$ cases proceed analogously.

In the $\fso(4j)$ case, there are no restrictions on brane wrappings, so an arbitrary bound state $[n_{\text{F1}},n_{\text{D1}},n_{\text{D5}},n_{\text{NS5}}]$ is possible. Two such bound states $$[m_{\text{F1}},m_{\text{D1}},m_{\text{D5}},m_{\text{NS5}}]\quad \text{and} \quad [n_{\text{F1}},n_{\text{D1}},n_{\text{D5}},n_{\text{NS5}}]$$ commute  if
\begin{equation}
n_{\text{F1}}m_{\text{NS5}}-n_{\text{NS5}}m_{\text{F1}}+n_{\text{D1}}m_{\text{D5}}-n_{\text{D5}}m_{\text{D1}}+(n_{\text{D5}}m_{\text{NS5}}-n_{\text{NS5}}m_{\text{D5}})j\in 2\mathbb{Z}. \label{eq.bulkboundcommutation}
\end{equation}
Motivated by the dictionary in table \ref{table:dictionary}, we define
\begin{align}
\begin{aligned}
(z_{es},z_{ec};z_{ms},z_{mc})&\df(n_{\text{F1}}+n_{\text{D5}},n_{\text{F1}};n_{\text{D1}}+n_{\text{NS5}},n_{\text{D1}})\,, \\
(z_{es}',z_{ec}';z_{ms}',z_{mc}')&\df(m_{\text{F1}}+m_{\text{D5}},m_{\text{F1}};m_{\text{D1}}+m_{\text{NS5}},m_{\text{D1}})\,.
\end{aligned}\label{eq.boundcharges}
\end{align}
In these new variables, the condition \eqref{eq.bulkboundcommutation} becomes
\begin{multline}
z_{ec}(z_{ms}'-z_{mc}')-(z_{ms}-z_{mc})z_{ec}'+z_{mc}(z_{es}'-z_{ec}')-(z_{es}-z_{ec})z_{mc}'\\+j((z_{es}-z_{ec})(z_{ms}'-z_{mc}')-(z_{ms}-z_{mc})(z_{es}'-z_{ec}'))\\
=\left(1+j\right)\left(z_{ec}z_{ms}-z_{es}z_{mc}'+z_{mc}z_{es}'-z_{ms}z_{ec}'\right)+\\
+j\left(z_{ec}z_{mc}'+z_{es}z_{ms}'-z_{mc}z_{ec}'-z_{ms}z_{es}'\right)\in 2\mathbb{Z}.\label{eq.boundmutuallocality}
\end{multline}
That is, the bulk-brane non-commutativity is equivalent to the mutual locality conditions of \cite{Aharony:2013hda}.

Note that the mutual locality conditions in \eqref{eq.boundmutuallocality} distinguishes $\fso(8l)$ from $\fso(8l+4)$. For instance, in the $\fso(8l)$ case, we have $j\in2\mathbb{Z}$, and so the commuting condition reduces to
\begin{equation*}
z_{ec}z_{ms}-z_{es}z_{mc}'+z_{mc}z_{es}'-z_{ms}z_{ec}'\in2\bZ.
\end{equation*}

Extending this result to $\fso(4k+2)$, $\fso(2n+1)$, and $\fsp(2n)$
theories is straightforward. The center of $\fso(4k+2)$ is
$\mathbb{Z}_4$, corresponding to the fact that condensing two D5 ``fat'' strings leaves behind an F1 string as explained by \cite{Witten:1998xy}, consistent with table
\ref{table:dictionary}. Analogously, condensing two NS5 fat strings leaves behind a D1 string. Thus we can write
the commutativity relation in terms of the number of lines from D5 and
NS5 branes, and we recover the $\mathbb{Z}_4\times\mathbb{Z}_4$
relation of \cite{Aharony:2013hda}.

In the case of $\fso(2n+1)$, the center is $\mathbb{Z}_2$ and only lines from F1, D1, and D5 bulk branes are present with the F1 lines being endable. This simplifies the commutativity relation to the $\mathbb{Z}_2\times\mathbb{Z}_2$ relation of \cite{Aharony:2013hda}. 
The $\fsp$ results are analogous.

\subsection{Interpreting the bulk effective action}
\label{sec.effectiveaction=bf}

For an alternate perspective, consider the effective action~\eqref{eqn:Stuckelbergaction}, generalizing \cite{Bergman:2022otk}:
\begin{multline}
  S = -2 \pi i \int_{\M{5}} \bigl[ n
  \bfB_{\text{F1}} \wedge d \bfB_{\text{D1}}+ 2 \bfB_{\text{F1}} \wedge
  (d \bfB_{\text{NS5}} - \theta_{\text{RR}}\bfA_3) \\ + 2 \bfB_{\text{D1}} \wedge (d \bfB_{\text{D5}}- \theta_{\text{NS}}\bfA_3) +2 \bfA_1 \wedge d \bfA_3 \bigr] \, .
\end{multline}
When $\theta_{\text{RR}}=0=\theta_{\text{NS}}$ and $n\in2\bZ$ we can perform a $\text{GL}(4,\bZ)$ field redefinition to obtain
\begin{equation}
S=-2\pi i\int_{\M{5}} \bigl[2\bfB_2\wedge d\bfC_2+2\tilde{\bfB}_2\wedge d\tilde{\bfC}_2+2\bfA_1\wedge d\bfA_3 \bigr].
\end{equation}
This describes a $\bZ_2$ 0-form bulk gauge theory and a $\bZ_2\times\bZ_2$ 1-form bulk gauge theory, and these correspond to the global symmetries in the field theory side.

Meanwhile, when $\theta_{\text{RR}}=0=\theta_{\text{NS}}$ and $n\in2\bZ+1$ we can perform a $\text{GL}(4,\bZ)$ field redefinition to obtain
\begin{equation}
S=-2\pi i\int_{\M{5}} \bigl[ 4\bfB_2\wedge d\bfC_2+2\bfA_1\wedge d\bfA_3 \bigr].
\end{equation}
This describes a $\bZ_2$ 0-form bulk gauge theory and a $\bZ_4$ 1-form bulk gauge theory.

Finally, when $\theta_{\text{RR}}=\frac{1}{2}$ and $\theta_{\text{NS}}=0$ we, through a St\"uckelberg mechanism, integrate out $\bfA_1$ to obtain
\begin{equation}
S=-2\pi i\int_{\M{5}}2\bfB_{\text{D1}}\wedge d\bfB_{\text{D5}}.
\end{equation}
This is a $\bZ_2$ 1-form gauge theory. The other two choices of discrete torsion can be obtained from this one via $\SL{2,\mathbb Z}$ transformations.

\subsection{\alt{$\SL{2,\mathbb Z}$}{SL(2,Z)} duality webs}\label{sec.k=2-orbits}

In this section, we display the $\SL{2,\mathbb Z}$ duality webs of all
of the $k=2$ theories. As in \cite{Aharony:2013hda}, the possible theories are classified by their line
operator content, which we specify by listing generators for the lines of that theory. Meanwhile, through
the dictionary in table \ref{table:dictionary}, each theory is also
classified by the boundary conditions for bulk strings / fat strings. These two
different classifications are equivalent, and here we demonstrate that
the classifications are consistent under the action of
$\SL{2,\mathbb Z}$. That is, the duality webs of $\SL{2,\mathbb Z}$ on
both the bulk description and the field theory description are
equivalent.

\subsubsection*{\underline{$\fso(4j)$}}

As is evident in \S\ref{sec.Sfolds-fso(4j)}, and also discussed in \cite{Aharony:2013hda}, the mutual locality conditions for $\fso(8j)$ and $\fso(8j+4)$ differ. However, there are some duality webs that are common to both cases. We display these below, followed by the remaining duality webs which differ between the $\fso(8j)$ and $\fso(8j+4)$ cases.

\begin{figurehere}
\begin{tikzcd}
	\begin{pmatrix}\stackrel{\text{F1}}{(1,1;0,0)},\ \stackrel{\text{D5}}{(1,0;0,0)}\\\Spin(4j)\end{pmatrix}\arrow[loop left,looseness=3]{}{\text{T}}\arrow[d,"\text{S}"]
	&\hspace{-0.5cm}\begin{pmatrix}\stackrel{\text{F1}}{(1,1;0,0)},\ \stackrel{\text{D1}+\text{D5}}{(1,0;1,1)}\\\SO{4j}_-\end{pmatrix}\arrow[loop right,looseness=3]{}{\text{T}}\arrow[d,"\text{S}"]\\
	\begin{pmatrix}\stackrel{\text{D1}}{(0,0;1,1)},\ \stackrel{\text{NS5}}{(0,0;1,0)}\\(\SO{4j}/\mathbb{Z}_2)_{\begin{smallmatrix}+&+\\+&+\end{smallmatrix}}\end{pmatrix}\arrow[u]\arrow[d,"\text{T}"]
	&\hspace{-0.5cm}\begin{pmatrix}\stackrel{\text{D1}}{(0,0;1,1)},\ \stackrel{\text{F1}+\text{NS5}}{(1,1;1,0)}\\(\SO{4j}/\mathbb{Z}_2)_{\begin{smallmatrix}-&-\\-&-\end{smallmatrix}}\end{pmatrix}\arrow[u]\arrow[d,"\text{T}"]\\
	\begin{pmatrix}\stackrel{\text{F1}+\text{D1}}{(1,1;1,1)},\ \stackrel{\text{NS5}+\text{D5}}{(1,0;1,0)}\\(\SO{4j}/\mathbb{Z}_2)_{\begin{smallmatrix}-&+\\+&-\end{smallmatrix}}\end{pmatrix}\arrow[u]\arrow[loop left,looseness=3]{}{\text{S}}
	&\hspace{-0.5cm}\begin{pmatrix}\stackrel{\text{F1}+\text{D1}}{(1,1;1,1)},\ \stackrel{\text{F1}+\text{NS5}+\text{D5}}{(0,1;1,0)}\\(\SO{4j}/\mathbb{Z}_2)_{\begin{smallmatrix}+&-\\-&+\end{smallmatrix}}\end{pmatrix}\arrow[u]\arrow[loop right,looseness=3]{}{\text{S}}
\end{tikzcd}
\begin{tikzcd}
	\begin{pmatrix}\stackrel{\text{F1}}{(1,1;0,0)},\ \stackrel{\text{D1}}{(0,0;1,1)}\\\SO{4j}_+\end{pmatrix}\arrow[loop left,looseness=3]{}{\text{S,T}}
\end{tikzcd}
\captionof{figure}{The $\SL{2,\mathbb Z}$ webs shared by both $\fso(8j)$ and $\fso(8j+4)$ theories.}
\end{figurehere}

\vspace{5pt}

\begin{figurehere}
\begin{tikzcd}
	\begin{pmatrix}\stackrel{\text{F1}+\text{D5}}{(0,1;0,0)},\ \stackrel{\text{F1}+\text{D1}+\text{NS5}}{(1,1;0,1)}\\\text{Sc}(8j)_-\end{pmatrix}\arrow[loop left,looseness=3]{}{\text{T}}\arrow[d,"\text{S}"]
	&\hspace{-.5cm}\begin{pmatrix}\stackrel{\text{D5}}{(1,0;0,0)},\ \stackrel{\text{F1}+\text{NS5}}{(1,1;1,0)}\\\text{Ss}(8j)_-\end{pmatrix}\arrow[loop right,looseness=3]{}{\text{T}}\arrow[d,"\text{S}"]\\
	\begin{pmatrix}\stackrel{\text{D1}+\text{NS5}}{(0,0;0,1)},\ \stackrel{\text{F1}+\text{D1}+\text{D5}}{(0,1;1,1)}\\(\SO{8j}/\mathbb{Z}_2)_{\begin{smallmatrix}+&-\\+&+\end{smallmatrix}}\end{pmatrix}\arrow[u]\arrow[d,"\text{T}"]
	&\hspace{-.5cm}\begin{pmatrix}\stackrel{\text{NS5}}{(0,0;1,0)},\ \stackrel{\text{D1}+\text{D5}}{(1,0;1,1)}\\(\SO{8j}/\mathbb{Z}_2)_{\begin{smallmatrix}+&+\\-&+\end{smallmatrix}}\end{pmatrix}\arrow[u]\arrow[d,"\text{T}"]\\
	\begin{pmatrix}\stackrel{\text{F1}+\text{NS5}}{(1,1;1,0)},\ \stackrel{\text{D1}+\text{D5}}{(1,0;1,1)}\\(\SO{8j}/\mathbb{Z}_2)_{\begin{smallmatrix}-&-\\+&-\end{smallmatrix}}\end{pmatrix}\arrow[u]\arrow[loop left,looseness=3]{}{\text{S}}
	&\hspace{-.5cm}\begin{pmatrix}\stackrel{\text{NS5}+\text{D5}}{(1,0;1,0)},\ \stackrel{\text{F1}+\text{D1}+\text{D5}}{(0,1;1,1)}\\(\SO{8j}/\mathbb{Z}_2)_{\begin{smallmatrix}-&+\\-&-\end{smallmatrix}}\end{pmatrix}\arrow[u]\arrow[loop right,looseness=3]{}{\text{S}}\\
	\begin{pmatrix}\stackrel{\text{F1}+\text{D5}}{(0,1;0,0)},\ \stackrel{\text{D1}+\text{NS5}}{(0,0;0,1)}\\\text{Sc}(8j)_+\end{pmatrix}\arrow[loop left,looseness=3]{}{\text{S,T}}
	&\hspace{-.5cm}\begin{pmatrix}\stackrel{\text{D5}}{(1,0;0,0)},\ \stackrel{\text{NS5}}{(0,0;1,0)}\\\text{Ss}(8j)_+\end{pmatrix}\arrow[loop right,looseness=3]{}{\text{S,T}}
\end{tikzcd}
\captionof{figure}{$\SL{2,\mathbb Z}$ webs unique to $\fso(8j)$ theories.}
\end{figurehere}

\vspace{5pt}

\begin{figurehere}
\begin{tikzcd}
	\begin{pmatrix}\stackrel{\text{F1}+\text{D5}}{(0,1;0,0)},\ \stackrel{\text{NS5}}{(0,0;1,0)}\\\text{Sc}(8j+4)_+\end{pmatrix}\arrow[d,"\text{T}"]\arrow[r,"\text{S}"]&
	\begin{pmatrix}\stackrel{\text{D5}}{(1,0;0,0)},\ \stackrel{\text{D1}+\text{NS5}}{(0,0;0,1)}\\\text{Ss}(8j+4)_+\end{pmatrix}\arrow[d,"\text{T}"]\arrow[l]\\
	\begin{pmatrix}\stackrel{\text{F1}+\text{D5}}{(0,1;0,0)},\ \stackrel{\text{NS5}+\text{D5}}{(1,0;1,0)}\\(\text{Sc}(8j+4)_-\end{pmatrix}\arrow[u]\arrow[d,"\text{S}"]&
	\begin{pmatrix}\stackrel{\text{D5}}{(1,0;0,0)},\ \stackrel{\text{F1}+\text{D1}+\text{NS5}}{(1,1;0,1)}\\(\text{Ss}(8j+4)_-\end{pmatrix}\arrow[u]\arrow[d,"\text{S}"]\\
	\begin{pmatrix}\stackrel{\text{D1}+\text{NS5}}{(0,0;0,1)},\ \stackrel{\text{D1}+\text{D5}}{(1,0;1,1)}\\(\SO{8j+4}/\mathbb{Z}_2)_{\begin{smallmatrix}-&+\\+&+\end{smallmatrix}}\end{pmatrix}\arrow[u]\arrow[r,"\text{T}"]&
	\begin{pmatrix}\stackrel{\text{NS5}}{(0,0;1,0)},\ \stackrel{\text{F1}+\text{D1}+\text{D5}}{(0,1;1,1)}\\(\SO{8j+4}/\mathbb{Z}_2)_{\begin{smallmatrix}+&+\\+&-\end{smallmatrix}}\end{pmatrix}\arrow[u]\arrow[l]\\
	\begin{pmatrix}\stackrel{\text{D1}+\text{D5}}{(1,0;1,1)},\ \stackrel{\text{F1}+\text{D1}+\text{NS5}}{(1,1;0,1)}\\(\SO{8j+4}/\mathbb{Z}_2)_{\begin{smallmatrix}+&-\\-&-\end{smallmatrix}}\end{pmatrix}\arrow[r,"\text{S, T}"]&
	\begin{pmatrix}\stackrel{\text{F1}+\text{NS5}}{(1,1;1,0)},\ \stackrel{\text{F1}+\text{D1}+\text{D5}}{(0,1;1,1)}\\(\SO{8j+4}/\mathbb{Z}_2)_{\begin{smallmatrix}-&-\\-&+\end{smallmatrix}}\end{pmatrix}\arrow[l]
\end{tikzcd}
\captionof{figure}{$\SL{2,\mathbb Z}$ webs unique to $\fso(8j+4)$ theories.}
\end{figurehere}
These diagrams agree with figures 7 and 8 of \cite{Aharony:2013hda}.

\subsubsection*{\underline{$\fso(4j+2)$}}

To map out the duality web in this case, note that---unlike in the $\fso(4j)$ case---a pair of fat D5 strings leave behind an F1 string when they annihilate~\cite{Witten:1998xy}. Thus, repeatedly applying T takes $\text{NS5}\to\text{NS5}+\text{D5}\to\text{NS5}+2\text{D5}=\text{NS5}+\text{F1}$, etc.

As noted in table \ref{table:dictionary}, the relative sign of the D5 and NS5 brane charges under the maximal 1-form symmetry is not fixed by our analysis, so we keep it arbitrary in the figure below:

\begin{figurehere}
\begin{tikzcd}
&\begin{pmatrix}\stackrel{\text{F1}}{(2,0)},\ \stackrel{\text{D1}}{(0,2)}\\\SO{4j+2}_+\end{pmatrix}\arrow[loop right,looseness=3]{}{\text{T, S}}&\\&
\begin{pmatrix}\stackrel{\text{F1}}{(2,0)},\  \stackrel{\text{D5}}{(1,0)}\\\Spin(4j+2)\end{pmatrix}\arrow[d,"\text{S}"]\arrow[loop right,looseness=3]{}{\text{T}}&\\
&\begin{pmatrix}\stackrel{\text{D1}}{(0,2)},\  \stackrel{\text{NS5}}{(0,\pm1)}\\(\Spin(4j+2)/\mathbb Z_4)_0\end{pmatrix}\arrow[u]\arrow[ld,"\text{T}"]&\\
\begin{pmatrix}\stackrel{\text{F1}+\text{D1}}{(2,2)},\ \stackrel{\text{NS5}+\text{D5}}{(1,\pm1)}\\(\Spin(4j+2)/\mathbb Z_4)_{\pm1}\end{pmatrix}\arrow[rd,"\text{T}"]\arrow[rr,"\text{S}"]&
&\begin{pmatrix}\stackrel{\text{F1}+\text{D1}}{(2,2)},\  \stackrel{\overline{\text{NS5}}+\text{D5}}{(1,\mp1)}\\(\Spin(4j+2)/\mathbb Z_4)_{\mp1}\end{pmatrix}\arrow[lu,"\text{T}"]\arrow[ll]\\
&\begin{pmatrix}\stackrel{\text{D1}}{(0,2)},\  \stackrel{\text{F1}+\text{NS5}}{(2,\pm1)}\\(\Spin(4j+2)/\mathbb Z_4)_2\end{pmatrix}\arrow[ru,"\text{T}"] \arrow[d,"\text{S}"]&\\
&\begin{pmatrix}\stackrel{\text{F1}}{(2,0)},\  \stackrel{\text{D1}+\text{D5}}{(1,2)}\\\SO{4j+2}_-\end{pmatrix}\arrow[u]\arrow[loop right,looseness=3]{}{\text{T}}
\end{tikzcd}
\captionof{figure}{$\SL{2,\mathbb Z}$ webs for $\fso(4j+2)$ theories.}
\end{figurehere}

Since the relative sign of the NS5 and D5 center charge affects the labels of $(\Spin(4j+2)/\bZ_4)_{\pm1}$, by comparing with \cite{Aharony:2013hda} we can deduce that the two have center charges of the same sign for $\fso(8k+6)$ and of opposite sign for $\fso(8k+2)$. This suggests that the choice of quadratic refinement is sensitive to the amount of 5-form flux supporting the geometry, but we have not attempted to derive this fact from string theory. Up to this subtlety, this diagram agrees with figure 6 of \cite{Aharony:2013hda}.

\subsubsection*{\underline{$\fso(2n+1)$ and $\fsp(2n)$}}

There are a few extra subtleties in this case, which is why we have saved it for last. Firstly, note that in our paper, S and T refer to the generators of the $SL(2,\mathbb{Z})$ self-duality of type IIB string theory. As a consequence, our T generator differs slightly from the identically-named field theory operation defined in \cite{Aharony:2013hda}, which we denote $\widehat{\text{T}}$. We have ignored this distinction so far because $\widehat{\text{T}} = \text{T}$ in most cases, but for $\fsp(2n)$ theories $\widehat{\text{T}} = \text{T}^2$, related to the fact that the O3$^+$ is not T invariant~\cite{Witten:1998xy}.\footnote{One also finds $\widehat{\text{T}} = \text{T}^{1/2}$ for $\fso(3)$, but for simplicity we will ignore this low-$n$ special case .}

A second subtlety relates to the action of $\widehat{\text{T}} = \text{T}^2$ on the branes. As before, $\text{T}^2$ maps $\text{NS5}\to\text{NS5}+2\text{D5}$, so the result depends on the end product of the two D5 fat strings annihilating each other (but with the added subtlety that individual D5 fat strings cannot be isolated due to the $B_2$ torsion). By analogy with before we expect that either complete annihilation occurs or an F1 string is left behind, depending on whether $n$ is even or odd, respectively. Assuming this to be true, we obtain the following duality webs:

\begin{figurehere}
\begin{tikzcd}
	\begin{pmatrix}\stackrel{\text{F1}}{(0,0)},\  \stackrel{\text{D1}}{(0,1)}\\\SO{2n+1}_+\end{pmatrix}\arrow[loop left,looseness=3]{}{\text{T}}\arrow[r,"\text{S}"]
	&\begin{pmatrix}\stackrel{\text{D1}}{(0,0)},\ \stackrel{\text{F1}}{(1,0)}\\\Sp{2n}\end{pmatrix}\arrow[l]\arrow[loop right,looseness=3]{}{\text{T}^2}\\
	\begin{pmatrix}\stackrel{\text{F1}}{(0,0)},\ \stackrel{\text{D5}}{(1,0)}\\\Spin(2n+1)\end{pmatrix}\arrow[loop left,looseness=3]{}{\text{T}}\arrow[r,"\text{S}"]&
	\begin{pmatrix}\stackrel{\text{D1}}{(0,0)},\  \stackrel{\text{NS5}}{(0,1)}\\(\Sp{2n}/\bZ_2)_+\end{pmatrix}\arrow[l]\arrow[loop right,looseness=3]{}{\text{T}^2}\\
	\begin{pmatrix}\stackrel{\text{F1}}{(0,0)},\  \stackrel{\text{D1+D5}}{(1,1)}\\\SO{2n+1}_-\end{pmatrix}\arrow[r,"\text{S}"]\arrow[loop left,looseness=3]{}{\text{T}}&
	\begin{pmatrix}\stackrel{\text{D1}}{(0,0)},\ \stackrel{\text{F1+NS5}}{(1,1)}\\(\Sp{2n}/\bZ_2)_-\end{pmatrix}\arrow[loop right,looseness=3]{}{\text{T}^2}\arrow[l]
\end{tikzcd}
\captionof{figure}{$\SL{2,\mathbb Z}$ webs for $\fso(2n+1)$ and $\fsp(2n)$ theories with even $n$.}
\end{figurehere}

\vspace{5pt}

\begin{figurehere}
\begin{tikzcd}
	\begin{pmatrix}\stackrel{\text{F1}}{(0,0)},\  \stackrel{\text{D1}}{(0,1)}\\\SO{2n+1}_+\end{pmatrix}\arrow[loop left,looseness=3]{}{\text{T}}\arrow[r,"\text{S}"]
	&\begin{pmatrix}\stackrel{\text{D1}}{(0,0)},\ \stackrel{\text{F1}}{(1,0)}\\\Sp{2n}\end{pmatrix}\arrow[l]\arrow[loop right,looseness=3]{}{\text{T}^2}\\
	\begin{pmatrix}\stackrel{\text{F1}}{(0,0)},\ \stackrel{\text{D5}}{(1,0)}\\\Spin(2n+1)\end{pmatrix}\arrow[loop left,looseness=3]{}{\text{T}}\arrow[r,"\text{S}"]&
	\begin{pmatrix}\stackrel{\text{D1}}{(0,0)},\  \stackrel{\text{NS5}}{(0,1)}\\(\Sp{2n}/\bZ_2)_+\end{pmatrix}\arrow[l]\arrow[d]\\
	\begin{pmatrix}\stackrel{\text{F1}}{(0,0)},\  \stackrel{\text{D1+D5}}{(1,1)}\\\SO{2n+1}_-\end{pmatrix}\arrow[r,"\text{S}"]\arrow[loop left,looseness=3]{}{\text{T}}&
	\begin{pmatrix}\stackrel{\text{D1}}{(0,0)},\ \stackrel{\text{F1+NS5}}{(1,1)}\\(\Sp{2n}/\bZ_2)_-\end{pmatrix}\arrow[u,"\text{T}^2"]\arrow[l]
\end{tikzcd}
\captionof{figure}{$\SL{2,\mathbb Z}$ webs for $\fso(2n+1)$ and $\fsp(2n)$ theories with odd $n>1$.}
\end{figurehere}

The agreement between these figures and figure 5 of \cite{Aharony:2013hda} validates our guess about the end product of the annihilation of two D5 fat strings. It would interesting to derive this directly from string theory. 

\section{Conclusions} \label{sec:conclusions}

In this paper we have developed a holographic description of the
symmetry operators of $\cN=3$ and $\cN=4$ 4d SCFTs via S-folds. The
$\cN=4$ S-folds are dual to SYM theories with BCD gauge algebras, and
in this setting our results for the symmetry operators, anomaly,
global forms and $\SL{2,\mathbb{Z}}$ orbits are consistent with
previous literature. The $\cN=3$ S-folds are dual to non-Lagrangian
SCFTs, and our analysis provides novel data on their symmetries.

There is an aspect of our analysis that was not fully justified, that
we would like to highlight: in the derivation of the Freed-Witten
anomaly cancellation condition, we assumed that there was no
non-perturbative contribution on the right hand side
of~\eqref{eq.Freed-Witten-linking} (akin to the $W_3$ term appearing
in \cite{Freed:1999vc}). Although we gave some circumstantial evidence
for the validity of our assumption, it would be interesting to verify
if our assumption is valid by a direct analysis of the M5 brane
anomaly
\cite{Witten:1999vg,Diaconescu:2003bm,Belov:2006jd,Monnier:2013kna,Monnier:2013rpa}.

Even in the $\cN=4$ case where the field theory dual is well
understood, our analysis is not quite complete. In the theories dual
to $\fso(4N+2)$ there are two inequivalent choices for the quadratic
refinement. The choice which correctly reproduces the full
$\SL{2,\bZ}$ orbits depends on $N$. In the field theory this
difference is related to the fractional instanton number (see for
instance \cite{Witten:2000nv,Aharony:2013hda,Cordova:2019uob} for the
computation of fractional instanton numbers using field theory
methods), so we expect that a careful treatment of the relevant
quadratic refinement of the string theory action should reproduce this
dependence on the fractional instanton number, as in the examples
considered in \cite{GarciaEtxebarria:2019caf,Apruzzi:2021nmk}.

The techniques developed in this paper should be extendable to $\cN<3$
S-folds (see for instance \cite{Apruzzi:2020pmv,Giacomelli:2020jel}
for pioneering work in this direction), and a geometric
characterization of the symmetries of those theories would be
interesting. An alternative direction for generalization is the class
of $\cN=1$ orientifold SCFTs studied in
\cite{Garcia-Etxebarria:2012ypj,Garcia-Etxebarria:2013tba,Garcia-Etxebarria:2015hua,Garcia-Etxebarria:2016bpb}.

Another direction for further study would be to derive the fusion
rules for symmetry generators in the non-invertible case directly
using D-brane methods. (See \cite{Apruzzi:2022rei} for a study of this
problem in a different system.) The non-trivial duality bundle on the
non-abelian theory on the stack of D3 branes --- or alternatively, the
poorly understood dynamics of the non-abelian $(2,0)$ theory in six
dimensions --- should make this computation fairly interesting.

\acknowledgments

We are very thankful to Mathew Bullimore, Jonathan Heckman, Saghar
Hosseini, Craig Lawrie, Fernando Marchesano, Diego Regalado, Sakura
Sch\"afer-Nameki, Yuji Tachikawa and Gabi Zafrir for illuminating
discussions and comments. The work of ME, BH, and SR was supported by
NSF grants PHY-1914934 and PHY-2112800. IGE is partially supported by
STFC consolidated grant ST/T000708/1 and the Simons
Collaboration on Global Categorical Symmetries.

\bibliographystyle{JHEP}
\bibliography{refs}
\end{document}